\documentclass[review,12pt]{elsarticle}

\usepackage{amssymb}
\usepackage{amsthm}
\usepackage{amsmath}

\usepackage{mathrsfs}
\usepackage{graphicx}
\usepackage{epstopdf}
\usepackage{float}
\usepackage{caption}
\usepackage{subcaption}
\usepackage{bm}
\usepackage{bbm}
\usepackage{mathrsfs}
\usepackage{cleveref}
\usepackage{soul}
\usepackage{accents}
\usepackage{graphicx}
\usepackage{xcolor}
\usepackage{courier} 
\usepackage{listings} 
\usepackage{tabu} 
\usepackage{longtable}
\usepackage{changepage} 
\usepackage{tabularx}
\usepackage{upgreek} 
\biboptions{sort&compress} 

\journal{CMAME}

\makeatletter
\def\@author#1{\g@addto@macro\elsauthors{\normalsize%
    \def\baselinestretch{1}%
    \upshape\authorsep#1\unskip\textsuperscript{%
      \ifx\@fnmark\@empty\else\unskip\sep\@fnmark\let\sep=,\fi
      \ifx\@corref\@empty\else\unskip\sep\@corref\let\sep=,\fi
      }%
    \def\authorsep{\unskip,\space}%
    \global\let\@fnmark\@empty
    \global\let\@corref\@empty  
    \global\let\sep\@empty}%
    \@eadauthor={#1}
}
\makeatother

\begin{document}

\begin{frontmatter}



\title{A generalised phase field model for fatigue crack growth in elastic-plastic solids with an efficient monolithic solver}


\author[IC]{Zeyad Khalil}

\author[IC]{Ahmed Y. Elghazouli}

\author[IC]{Emilio Mart\'{\i}nez-Pa\~neda\corref{cor1}}
\ead{e.martinez-paneda@imperial.ac.uk}

\address[IC]{Department of Civil and Environmental Engineering, Imperial College London, London SW7 2AZ, UK}

\cortext[cor1]{Corresponding author.}

\begin{abstract}
We present a generalised phase field-based formulation for predicting fatigue crack growth in metals. The theoretical framework aims at covering a wide range of material behaviour. Different fatigue degradation functions are considered and their influence is benchmarked against experiments. The phase field constitutive theory accommodates the so-called \texttt{AT1}, \texttt{AT2} and phase field-cohesive zone (\texttt{PF-CZM}) models. In regards to material deformation, both non-linear kinematic and isotropic hardening are considered, as well as the combination of the two. Moreover, a monolithic solution scheme based on quasi-Newton algorithms is presented and shown to significantly outperform staggered approaches. The potential of the computational framework is demonstrated by investigating several 2D and 3D boundary value problems of particular interest. Constitutive and numerical choices are compared and insight is gained into their differences and similarities. The framework enables predicting fatigue crack growth in arbitrary geometries and for materials exhibiting complex (cyclic) deformation and damage responses. The finite element code developed is made freely available at www.empaneda.com/codes.
\end{abstract}


\end{frontmatter}


\section{Introduction}
\label{Introduction}

Fatigue-induced fracture is considered to be one of the most frequent causes of failure in engineering components. Due to its complex nature, the development of computational models capable of predicting fatigue cracking is regarded to be highly challenging and has been the subject of extensive research. Generally, the evolution of fatigue damage occurs in different stages. Firstly, nucleation of permanent damage takes place as a result of sub- and microstructural changes within the material, followed by the creation of microscopic cracks. Subsequently, these microscopic flaws start growing and eventually coalescence, leading to the formation of dominant macro-cracks. These macro-cracks then propagate leading to structural instability and complete fracture of the component. Fatigue design is commonly based on classical empirical methods \cite{Suresh1998}. Such methods involve the estimation of the total life to failure as a function of the cyclic stress range, which is referred to as the S-N curve approach, or the strain range (plastic or total). These fatigue design strategies are commonly referred to as total-life approaches, where the fatigue life is defined as the number of cycles ($N_f$) or reversals ($2N_f$) to failure. The stress-life or S-N curve approach was first developed by W\"{o}hler \cite{Wohler1870}. Under low cyclic stress amplitudes, the material behaves mainly in an elastic manner and a very large number of cycles are required to cause failure; normally more than 10$^6$, which is referred to as High-cycle Fatigue (HCF). This approach has become popular in applications involving low-amplitude cyclic stresses such as steel bridges under traffic loading and railway axles. On the other hand, a much lower number of cycles are needed to cause failure if the applied stresses are large enough to cause plastic deformations; in the order of 10$^2$ to 10$^4$, a regime referred to as Low-cycle Fatigue (LCF). In such situations, fatigue life is determined in terms of the applied strain range, as first proposed by Coffin \cite{Coffin1954} and Manson \cite{Manson1954}. In the case of components experiencing very large plastic deformations, complete structural failure can occur after very few cycles \cite{Nip2010,Nip2010a}. This is often referred to as Extremely- or Ultra-low Cycle fatigue (ELCF or ULCF). Due to their empirical nature, stress-life and strain-based approaches have limited applicability and cannot be readily generalised to arbitrary materials, geometries and loading histories.\\ 

Variational phase field fracture models can provide a reliable computational framework to predict High-, Low- and Extremely low-cycle fatigue, overcoming the limitations of semi-empirical approaches. Phase field fracture methods have been gaining increasing attention. Predictions are based on the thermodynamic framework outlined in the works of Griffith \cite{Griffith1920} and Irwin \cite{Irwin1956}, where a crack would only propagate if the energy release rate exceeds a critical value, the material toughness $G_c$. Francfort and Marigo \cite{Francfort1998} introduced a variational formulation for Griffith's thermodynamical framework, including the surface energy dissipated due to crack formation in the total potential energy. Inspired by the work of Mumford and Shah \cite{mumford1989} on image segmentation, Bourdin \textit{et al.} \cite{Bourdin2000,Bourdin2008} introduced a scalar phase field variable that regularises the discrete crack topology into a diffuse crack representation. In addition, Miehe and co-workers \cite{Miehe2010,Miehe2010a} have made significant contributions to the development of the method by proposing new robust solution schemes.\\ 

Phase field fracture methods have been used in numerous applications, including dynamic fracture \cite{Borden2012,Geelen2019,Molnar2020}, cracking of composite materials \cite{Alessi2019,Quintanas-Corominas2020a,Mandal2020,CST2021}, ductile damage \cite{Ambati2015a,Borden2016,IJP2021}, hydrogen-assisted cracking \cite{CMAME2018,Duda2018,Wu2020b,TAFM2020c}, and fracture of functionally graded materials \cite{CPB2019,Kumar2021}, among many others; see Refs. \cite{Wu2020,PTRSA2021} for an overview. Recently, the success of variational phase field methods has been extended to fatigue crack growth \cite{Alessi2018c,Lo2019,Carrara2020,CMAME2021,Hasan2021}. This is an exciting and natural extension for phase field fracture; Griffith's thermodynamic principles should hold for fatigue crack growth and metal fatigue was actually the motivation for Griffith's seminal work \cite{Griffith1920}. Lo \textit{et al.} \cite{Lo2019} combined a phase field kinetic law with a modified $J$-integral to capture Paris-law type fatigue crack growth behaviour. Carrara \textit{et al.} \cite{Carrara2020} proposed a novel variational framework to capture the fatigue behaviour of brittle materials by introducing a fatigue degradation function that degrades the material toughness. Simoes and Mart\'{\i}nez-Pa\~neda \cite{CMAME2021} simulated the fatigue failure of a NiTi stent, accounting for both fatigue damage and phase transformations. Hasan and Baxevanis \cite{Hasan2021} introduced a toughness degradation law through a measure of accumulated elastic strain energy density that enabled capturing both total life and defect tolerant approaches to fatigue. Golahmar \textit{et al.} presented a phase field formulation for hydrogen assisted fatigue \cite{Golahmar2022}. However, these studies are all limited to the analysis of linear elastic materials; only very recently have phase field methods been used to predict fatigue crack growth in elastic-plastic solids. Seiler \textit{et al.} \cite{Seiler2020} used a local strain approach to empirically incorporate plasticity \textit{via} Neuber's rule. Haveroth \textit{et al.} \cite{Haveroth2020} presented a new thermo-mechanical fatigue formulation using a Voce-type hardening law and a new degradation function that degrades both elastic and plastic strain energy densities. Finally, Ulloa \textit{et al.} \cite{Ulloa2021} developed a phase field fatigue formulation for elastic-plastic solids suitable for both low and high cycle fatigue regimes and capable of capturing ratcheting effects.\\

In this work, we present a generalised formulation for fatigue damage in metals. We aim to model a general class of elastic-plastic materials and thus account for the combination of non-linear isotropic and kinematic hardening effects. Moreover, unlike previous work, we do not restrict our attention to one class of phase field fracture models but accommodate both brittle and quasi-brittle formulations; namely, the so-called \texttt{AT1}, \texttt{AT2} and \texttt{PF-CZM} models. In addition, we couple our phase field fatigue framework with a quasi-Newton monolithic solution scheme and show that it is more robust and significantly more efficient than the widely-used staggered schemes. This is of notable importance given the computational cost associated with cycle-by-cycle fatigue predictions. Several 2D and 3D boundary value problems are investigated to gain insight into the various constitutive choices of the model for deformation and fracture. Firstly, we simulate the fatigue failure of planar specimens under uniaxial cyclic loading. Predictions are compared with experiments on a carbon steel that exhibits combined non-linear isotropic and kinematic hardening. Secondly, fatigue crack growth in a Compact Tension (CT) sample is investigated, evaluating the differences between various phase field fracture models and solution schemes. Thirdly, the failure of an asymmetrically-notched bar is simulated to gain insight into the interplay between hardening mechanisms and damage. Finally, we demonstrate the capabilities of the computational framework in predicting the failure of 3D components by modelling crack nucleation and growth in a pipe-to-pipe connection.\\

The remainder of this manuscript is organised as follows. The generalised theoretical framework presented is described in Section \ref{Sec:Theory}. In Section \ref{Sec:FEM} we provide details of the numerical implementation, including the monolithic quasi-Newton solution scheme. The results computed are shown and discussed in Section \ref{Sec:Results}. Finally, the manuscript ends with concluding remarks in Section \ref{Sec:Concluding remarks}.

\section{Theory}
\label{Sec:Theory}

In this section, we present our generalised formulation, suitable for arbitrary constitutive choices of crack density function, fracture driving force, degradation function and cyclic material response. The theory refers to an elastic-plastic body occupying an arbitrary domain $\Omega \subset {\rm I\!R}^n$ $(n \in[1,2,3])$, with an external boundary $\partial \Omega\subset {\rm I\!R}^{n-1}$, on which the outwards unit normal is denoted as $\mathbf{n}$. We shall first define the kinematic variables (Section \ref{Sec:Kinematics}), then derive the force balances using the principle of virtual power (Section \ref{Sec:PVW}), and finally particularise our theory to relevant constitutive choices for the deformation and fracture behaviour of the solid (Section \ref{Sec:ConstitutiveChoices}). 

\subsection{Kinematics}
\label{Sec:Kinematics}

The primary kinematic variables are the displacement field vector $\mathbf{u}$ and the damage phase field $\phi$. Small displacements and strains are assumed, such that the total strain tensor $\bm{\varepsilon}$ reads
\begin{equation}
    \bm{\varepsilon} = \frac{1}{2}\left(\nabla\mathbf{u}^T+\nabla\mathbf{u}\right) \, .
\end{equation}

\noindent Also, we adopt the standard decomposition of strains into elastic and plastic components, such that:
\begin{equation}
    \bm{\varepsilon} = \bm{\varepsilon}^e+ \bm{\varepsilon}^p \, .
\end{equation}

The nucleation and growth of cracks are described by using a smooth continuous scalar \emph{phase field} $\phi \in [0;1]$. The use of an auxiliary phase field variable to \emph{implicitly} track interfaces has proven to be a very compelling computational approach for numerous interfacial problems, such as microstructural evolution \cite{Provatas2011} and metallic corrosion \cite{JMPS2021}. In the context of fracture mechanics, the phase field variable resembles a damage variable; it must grow monotonically $\dot{\phi} (\mathbf{x}, t) \geq 0$ and describes the degree of damage, with $\phi=1$ denoting fracture and $\phi=0$ corresponding to the intact phase. Since $\phi$ is smooth and continuous, discrete cracks are represented in a diffuse fashion, with the smearing of cracks being controlled by a phase field length scale $\ell$. The aim of this diffuse representation is to introduce, over a discontinuous surface $\Gamma$, the following approximation of the fracture energy \cite{Bourdin2000}:
\begin{equation}
    \Phi=\int_{\Gamma} G_c \, \text{d}S \approx \int_\Omega G_c\Upsilon(\phi,\nabla\phi) \, \text{d}V, \hspace{1cm} \text{for } \ell\rightarrow 0^+,
\end{equation}

\noindent where $\Upsilon$ is the so-called crack surface density functional and $G_c$ is the critical energy release rate or material toughness. We extend this rate-independent description of fracture to accommodate time and history dependent problems. Thus, for a cumulative history variable $\bar{\vartheta}$, which fulfills $\dot{\bar{\vartheta}}>0$, and a fatigue degradation function $f \left( \bar{\vartheta} \right)$, the fracture energy can be re-formulated as follows,
\begin{equation}\label{eq:Phi_new}
    \Phi= \int_0^t \left( \int_\Omega G_c f \left( \bar{\vartheta} \left(t \right) \right) \dot{\Upsilon}(\phi,\nabla\phi) \, \text{d} V \right) \, \text{d}t \, .
\end{equation}

As described below, this is complemented by appropriate constitutive choices that characterise the degradation of the fracture energy with the fatigue history variable $\bar{\vartheta}$. 

\subsection{Principle of virtual power. Balance of forces}
\label{Sec:PVW}

Now, let us derive the balance equations for the coupled deformation-fatigue system using the principle of virtual power. Define $\bm{\sigma}$ as the symmetric Cauchy tensor, $\mathbf{T}$ as a surface traction acting on the boundary of the solid $\partial\Omega$ and $\mathbf{b}$ as a prescribed body force per unit volume. With respect to the damage problem, we introduce a scalar stress-like quantity $\omega$, which is work conjugate to the phase field $\phi$, and a phase field micro-stress vector $\bm{\upxi}$, which is work conjugate to the gradient of the phase field $\nabla\phi$. No external traction is associated with the phase field. We define the kinematics of the body $\Omega$ using the fields $\mathbf{u}$ and $\phi$. The corresponding velocities read $\dot{\mathbf{u}}$ and $\dot{\phi}$, while the virtual velocities, defined over a vector space $\mathcal{V}$, are denoted by $\tilde{\mathbf{u}}$ and $\tilde{\phi}$ \cite{Gurtin2010,Duda2015,Narayan2019}. Accordingly, the principle of virtual power reads: 
\begin{equation}\label{eq:PVW}
 \int_\Omega \left[ \bm{\sigma}: \nabla \tilde{\mathbf{u}}  + \omega \tilde{\phi} +\bm{\upxi} \cdot \nabla \tilde{\phi}
   \right] \, \text{d}V =  \int_{\partial \Omega} \left( \mathbf{T} \cdot \tilde{\mathbf{u}}  \right) \, \text{d}S + \int_{\Omega} \left( \mathbf{b} \cdot \tilde{\mathbf{u}} \right) \, \text{d}V \, .
\end{equation}

By application of the Gauss divergence theorem and the fundamental lemma of calculus of variations, the local force balances are given by: 
\begin{equation}
    \begin{split}
        &\nabla\cdot\bm{\sigma} + \mathbf{b}=0  \\
        &\nabla\cdot\bm{\upxi}-\omega =0
    \end{split}\hspace{2cm} \text{in } \,\, \Omega,\label{eq:balance}
\end{equation}

\noindent with natural boundary conditions: 
\begin{equation}
    \begin{split}
        \bm{\sigma}\cdot\mathbf{n}=\mathbf{T} \\
         \bm{\upxi} \cdot \mathbf{n}=0 
    \end{split} \hspace{2cm} \text{on } \,\, \partial\Omega.\label{eq:balance_BC}
\end{equation}

\subsection{Constitutive theory}
\label{Sec:ConstitutiveChoices}

We shall now proceed to make suitable constitutive choices for the deformation, fracture and fatigue behaviour of the solid. First, we define the total potential energy of the solid as the sum of the strain energy density of the solid $\psi$ and the fracture energy density $\varphi$, such that:
\begin{equation}\label{eq:TotalPotentialEnergy0}
W \left( \bm{\varepsilon} \left( \mathbf{u} \right), \, \phi, \,  \nabla \phi \right) = \psi \left( \bm{\varepsilon} \left( \mathbf{u} \right), \, g \left( \phi \right) \right)  +  \varphi \left( \phi, \, \nabla \phi \right) \, ,
\end{equation}

\noindent where $g \left( \phi \right)$ is a phase field degradation function, to be defined. The Cauchy stress tensor is then defined as $\bm{\sigma}=\partial_{\bm{\varepsilon}} \psi$. Thus, the phase field reduces the stiffness of the solid. However, note that no damage-plasticity coupling term is defined. The reader is referred to Alessi \textit{et al.} \cite{Alessi2018} for a comprehensive discussion on potential constitutive choices to capture the interplay between damage and elastic-plastic material behaviour. The strain energy density includes both elastic and plastic parts, which are computed as follows:
\begin{equation}\label{eq:StrainEnergyDensity}
    \psi_{0} = \psi^e \left(\bm{\varepsilon}^e \right) + \psi^p \left( \bm{\varepsilon}^p \right)= \frac{1}{2} \lambda \left[ \text{tr} \left( \bm{\varepsilon}^e \right) \right]^2 + \mu \, \text{tr} \left[  \left( \bm{\varepsilon}^e \right)^2 \right] + \int_0^t \left( \bm{\sigma}_{0} : \dot{\bm{\varepsilon}}^p \right) \text{d} t \, ,
\end{equation}

\noindent where $\lambda$ and $\mu$ are the Lam\'{e} parameters and the subscript $0$ denotes an undamaged quantity. In agreement with (\ref{eq:Phi_new}), the fracture energy density is defined as, 
\begin{equation}
   \varphi = f \left( \bar{\vartheta}  \right) G_c \Upsilon(\phi, \nabla\phi) = f \left( \bar{\vartheta}  \right) \dfrac{G_c }{4c_w\ell}\left( w(\phi)  + \ell^2 | \nabla \phi |^2  \right) \, .
\end{equation}

\noindent Here, $w(\phi)$ is the geometric crack function, to be defined, and $c_w$ is a scaling constant, given by
\begin{equation}
    c_w = \int_0^1\sqrt{w(\zeta)} \, \text{d}\zeta \, .
\end{equation}

\subsubsection{Strain energy decomposition}

To prevent the nucleation and growth of cracks under compression, the strain energy density can be decomposed into tensile and compressive parts as follows:
\begin{equation}
    \psi \left( \bm{\varepsilon}^e, \bm{\varepsilon}^p, \phi \right) =g \left( \phi \right) \left( \psi_+^e \left( \bm{\varepsilon}^e \right)  + \psi^p \left( \bm{\varepsilon}^p \right) \right) + \psi^e_-  \left( \bm{\varepsilon}^e \right) \, .
\end{equation}

We choose to follow the so-called \emph{volumetric-deviatoric} split proposed by Amor \textit{et al.} \cite{Amor2009}. Accordingly, for a material with bulk modulus $K$, the elastic strain energy density is decomposed as,
\begin{equation}
    \psi^e_+ \left( \bm{\varepsilon} \right) = \frac{1}{2} K \langle \text{tr} \left( \bm{\varepsilon}^e \right) \rangle^2_+ + \mu \left( {\bm{\varepsilon}^e}' : {\bm{\varepsilon}^e}' \right) \, , \,\,\,\,\,\, \psi^e_- \left( \bm{\varepsilon}^e \right) = \frac{1}{2} K \langle \text{tr} \left( \bm{\varepsilon}^e \right) \rangle^2_- \, ,
\end{equation}

\noindent where ${\bm{\varepsilon}^e}'$ denotes the deviatoric part of the elastic strain tensor and $\langle \,\rangle$ are the Macaulay brackets. 

\subsubsection{Karush–Kuhn–Tucker (KKT) conditions}

Damage is an irreversible process and, as a consequence, the phase field evolution law must fulfill the condition $ \dot{\phi} \geq 0$. To achieve this, we follow Miehe and co-workers \cite{Miehe2010,Miehe2010a} and define a history variable field $\mathcal{H}$. Since the effective plastic work is assumed to increase monotonically, the history field variable only relates to the elastic fracture driving force, $\psi_+^e$. Thus, as dictated by the Karush–Kuhn–Tucker (KKT) conditions, the definition of $\mathcal{H}$ must satisfy:
\begin{equation}
     \psi_+^e - \mathcal{H} \leq 0 \text{,} \hspace{8mm} \dot{\mathcal{H}} \geq 0 \text{,} \hspace{8mm} \dot{\mathcal{H}}(\psi_+^e -\mathcal{H})=0 \, \, .
    \centering
\end{equation}

Accordingly, for a current time $t$, within a total time $t_t$, we define the history field as,
\begin{equation}\label{eq:HistoryField}
    \mathcal{H} = \text{max}_{t \in [0, t_t]} \psi_+^e \left( t \right) \, .
\end{equation}

\subsubsection{Fatigue damage}

The damage resulting from the application of cyclic loads is captured by means of a fatigue degradation function $f(\bar{\vartheta})$, a cumulative history variable $\bar{\vartheta}$, and a fatigue threshold parameter $\vartheta_T$. We follow the work by Carrara \textit{et al.} \cite{Carrara2020} on elastic solids and consider two fatigue degradation functions; one of an asymptotic form:
\begin{equation}\label{eq:fdeg1}
   {f}(\bar{\vartheta}(t))=
   \begin{cases}
            1 &         \text{if } \Bar{\vartheta}(t)\leq\vartheta_T\\
            \left(\frac{2\vartheta_T}{\Bar{\vartheta}(t)+\vartheta_T}\right)^2 &         \text{if } \Bar{\vartheta}(t)\geq\vartheta_T
   \end{cases}
\end{equation}

\noindent and a second one, of logarithmic form:
\begin{equation}\label{eq:fdeg2}
   {f}(\bar{\vartheta}(t))=
   \begin{cases}
            1 &         \text{if } \Bar{\vartheta}(t)\leq\vartheta_T\\
            \left[1-\kappa\text{log}\left(\frac{\Bar{\vartheta}(t)}{\vartheta_T}\right)\right]^2 &         \text{if } \vartheta_T\leq\Bar{\vartheta}(t)\leq\vartheta_T10^{1/\kappa}\\
            0 &         \text{if }\Bar{\vartheta}(t)\geq\vartheta_T10^{1/\kappa}
   \end{cases}
\end{equation}

\noindent where $\kappa$ is a material parameter that characterises the slope of the logarithmic function. The impact of these choices must be assessed against experiments. Here, we provide a comparison against testing data on hot-rolled structural steels (see Section \ref{Sec:Uniaxial}). In addition, the evolution of the fatigue history variable $\bar{\vartheta}$ is given by,
\begin{equation}\label{eq:alpha_bar1}
   \Bar{\vartheta}(t)=\int_0^t H(\vartheta\dot{\vartheta})|\dot{\vartheta}|\text{ d}t \, ,
\end{equation}

\noindent where $H(\vartheta\dot{\vartheta})$ is the Heaviside function such that $\bar{\vartheta}$ only evolves during loading.\\

Finally, it remains to define the fatigue threshold parameter $\vartheta_T$ and the fatigue history variable $\vartheta$. For the latter, we choose to adopt the following constitutive choice,
\begin{equation}
    \vartheta = g \left(\phi \right) \left( \mathcal{H} + \psi^p \right) \, ,
\end{equation}

\noindent which implies that fatigue damage is driven by both elastic and plastic straining. This choice is consistent with the definition of a fracture driving force driven by both elastic and plastic strain energy densities. However, note that the use of the history field (as opposed to $\psi_+^e$) aims at minimising the elastic contribution, as arguably appropriate in the context of low-cycle fatigue. For the fatigue threshold, we follow Carrara \textit{et al.} \cite{Carrara2020} and assume:
\begin{equation}\label{eq:threshold}
    \vartheta_T=\frac{G_c}{12 \ell} \, ,
\end{equation}

\noindent unless otherwise stated.

\subsubsection{Micro-force variables}

We proceed to derive, without loss of generality, the fracture micro-stress variables $\omega$ and $\bm{\upxi}$. First, considering (\ref{eq:TotalPotentialEnergy0}), (\ref{eq:StrainEnergyDensity}) and (\ref{eq:HistoryField}), we reformulate the total potential energy of the solid as,
\begin{equation}\label{eq:Free_energy}
     W =  {g(\phi)} \left( \mathcal{H} + \psi^p \right)  + f \left(\bar{\vartheta} \right) \frac{G_c}{4 c_w }  \left(\frac{w(\phi)}{ \ell}+\ell |\nabla \phi|^2\right) \, .
\end{equation}

Consequently, the scalar micro-stress $\omega$ is defined as:
\begin{equation}\label{eq:consOmega}
    \omega = \dfrac{\partial W}{\partial\phi} = {g^{\prime}(\phi)} \left( \mathcal{H} + \psi^p \right) +f \left(\bar{\vartheta} \right) \frac{G_c}{4c_w \ell}  w^{\prime}(\phi) \, ,
\end{equation}

\noindent and the phase field micro-stress vector $\bm{\upxi}$ reads,
\begin{equation}\label{eq:consXi}
    \bm{\upxi} = \dfrac{\partial W}{\partial\nabla\phi} = \frac{\ell}{2c_w} G_{\mathrm{c}} f \left(\bar{\vartheta} \right) \nabla \phi \, .
\end{equation}

The phase field evolution law (\ref{eq:balance}b) can be reformulated by taking into consideration the constitutive relations (\ref{eq:consOmega}) and (\ref{eq:consXi}), such that 
\begin{equation}\label{eq:PhaseFieldStrongForm}
  \frac{G_c f \left( \bar{\vartheta} \right)}{2c_w}  \left( \frac{w^{\prime}(\phi)}{2 \ell} -  \ell \nabla^2 \phi\right) - \frac{G_c \ell }{2c_w} \nabla \phi \nabla f \left( \bar{\vartheta}  \right)  + {g^{\prime}(\phi)} \left( \mathcal{H} + \psi^p \right) = 0  \, .
\end{equation}

It is evident from (\ref{eq:HistoryField}) and (\ref{eq:PhaseFieldStrongForm}) that the phase field evolution is driven by both elastic and plastic strain energy densities. This is in agreement with several phase field models for fracture in elastic-plastic solids \cite{Miehe2016b,Borden2016,Shishvan2021a}. However, other approaches have also been considered in the literature (see, e.g. \cite{Duda2015,JMPS2020}) and the choice is not straightforward. From a thermodynamical perspective, the majority of the energy stored in the solid (and thus available to facilitate crack growth) is elastic. However, failure in ductile fracture experiments is often driven by plastic phenomena and a simple energy balance is not suitable for crack growth processes involving significant plasticity; e.g., the assumption of an isothermal process is no longer valid as plastic flow constitutes a source of heat \cite{PTRSA2021}.

\subsubsection{Degradation and dissipation functions}

Now, we particularise our generalised framework by making specific choices for the fracture degradation function $g(\phi)$, the so-called dissipation function $w(\phi)$ \cite{Bleyer2018,Alessi2018c,Materials2021} and the fracture driving force threshold $\mathcal{H}_{min}$. Our theory captures both brittle and quasi-brittle phase field approaches, accommodating the so-called \texttt{AT1}, \texttt{AT2} and \texttt{PF-CZM} models. The \texttt{AT1} and \texttt{AT2} models are based on the Ambrosio and Tortorelli regularisaton of the Mumford-Shah functional \cite{Ambrosio1991,Bourdin2000}, with the former aimed at including a purely elastic response up to the onset of damage \cite{Pham2011}. The phase field cohesive zone model \texttt{PF-CZM} employed here is inspired by the work by Wu \cite{Wu2017} and Wu and Nguyen \cite{Wu2018a}, but it employs a fracture driving force based on the strain energy density, as in Ref. \cite{Geelen2019}. In other words, fracture is driven by both elastic and plastic strain energy densities in the three phase field models considered, as shown in (\ref{eq:PhaseFieldStrongForm}). \\ 

Firstly, we start by defining the phase field degradation function, which should satisfy: 
\begin{equation}
g \left( 0 \right) =1 , \,\,\,\,\,\,\,\,\,\, g \left( 1 \right) =0 , \,\,\,\,\,\,\,\,\,\, g' \left( \phi \right) \leq 0 \,\,\, \text{for} \,\,\, 0 \leq \phi \leq 1 \, .
\end{equation}

\noindent For the \texttt{AT1} and \texttt{AT2} models the same choice is adopted; a quadratic degradation function such that,
\begin{equation}\label{eq:gAT1AT2}
    g \left( \phi \right) = \left( 1 - \phi \right)^2  \, .
\end{equation}

\noindent While for the \texttt{PF-CZM} model, the following form is used,
\begin{equation}\label{eq:gPF-CZM}
   g \left( \phi \right) = \frac{\left( 1 - \phi \right)^2}{\left( 1 - \phi \right)^2 + m \phi \left( 1 - 0.5 \phi \right)} \,\,\,\,\,\,\, \text{with} \,\,\,\,\,\,\, m = \frac{3EG_c}{2 \ell \sigma_c^2} \, ,
\end{equation}

\noindent where $\sigma_c$ is the material strength.

Secondly, we define the dissipation function, which should fulfill the following conditions:
\begin{equation}
w \left( 0 \right) =0 , \,\,\,\,\,\,\,\,\,\, w \left( 1 \right) =1 , \,\,\,\,\,\,\,\,\,\, w' \left( \phi \right) \geq 0 \,\,\, \text{for} \,\,\, 0 \leq \phi \leq 1 \, .
\end{equation}

The specific choice $w(\phi)=\phi^2$ ($c=1/2$) recovers the \texttt{AT2} model while $w(\phi)=\phi$ ($c=2/3$) renders the \texttt{AT1} formulation. No threshold for damage exists in the \texttt{AT2} case as $w'(0)=0$, unlike in the \texttt{AT1} model. Finally, for the \texttt{PF-CZM} we have $w(\phi)=4\phi$ ($c_w=4/3$), where $w'(0) > 0$ - as in the \texttt{AT1} case. Accordingly, a damage threshold should be defined for both \texttt{AT1} and \texttt{PF-CZM} models; the following choices are adopted here,
\begin{equation}\label{eq:HminB}
   \texttt{AT1}: \,\,\, \mathcal{H}_{min} = \frac{3G_c}{16\ell} \, , \,\,\,\,\,\,\,\,\,\,\, \texttt{PF-CZM}: \,\,\, \mathcal{H}_{min} = \frac{\sigma_c^2}{2E}  \, .
\end{equation}

Accordingly, in the numerical implementation of the \texttt{AT1} and \texttt{PF-CZM} models, the history field is taken to be the maximum of (\ref{eq:HistoryField}) and $\mathcal{H}_{min}$.\\

It is also important to note that the \texttt{PF-CZM} model explicitly incorporates the material strength $\sigma_c$ into the constitutive equations - see (\ref{eq:gPF-CZM}) and (\ref{eq:HminB})b. For the \texttt{AT1} and \texttt{AT2}, a relation between the strength and the phase field length scale can be derived by considering the homogeneous solution to the phase field evolution law (see, e.g. Ref. \cite{PTRSA2021}):
\begin{equation}\label{eq:AT1AT2strength}
   \texttt{AT1}: \,\,\, \sigma_c = \sqrt{\frac{3EG_c}{8 \ell}} \, , \,\,\,\,\,\,\,\,\,\,\, \texttt{AT2}: \,\,\, \sigma_c = \frac{3}{16}\sqrt{\frac{3EG_c}{\ell}}  \, .
\end{equation}

\subsubsection{Cyclic deformation: combined isotropic/kinematic hardening}
\label{SubSec:Cyclic}

The constitutive choices for relating the stresses to the strains aim at modelling a general class of elastic-plastic materials. Specifically, a nonlinear combined isotropic/kinematic hardening model is used to capture a wide range of cyclic plasticity phenomena. The model is based on the work by Lemaitre and Chaboche \cite{Lemaitre1990}, where a von Mises yield criterion is combined with a non-linear description of isotropic and kinematic hardening effects.\\

For a material with current yield stress $\sigma_Y$, experiencing a deviatoric backstress $\bm{\alpha}'$, the assumed yield function reads,
\begin{equation}
    \mathcal{F} = \sqrt{\frac{2}{3} \left( \bm{\sigma}' - \bm{\alpha}' \right) : \left( \bm{\sigma}' - \bm{\alpha}' \right)} - \sigma_Y = 0 \, ,
\end{equation}

\noindent where the first term is the von Mises effective stress. The flow rule then reads,
\begin{equation}
    \dot{\bm{\varepsilon}}^p = \dot{\varepsilon}^p \frac{\partial \mathcal{F} }{\partial \bm{\sigma}}
\end{equation}

\noindent where $\dot{\varepsilon}^p$ is the equivalent plastic strain rate, defined as $\dot{\varepsilon}^p=\sqrt{ (2/3) \dot{\bm{\varepsilon}}^p :\dot{\bm{\varepsilon}}^p}$.\\ 

\begin{figure}[H]
     \centering
         \centering
         \includegraphics[width=0.9\textwidth]{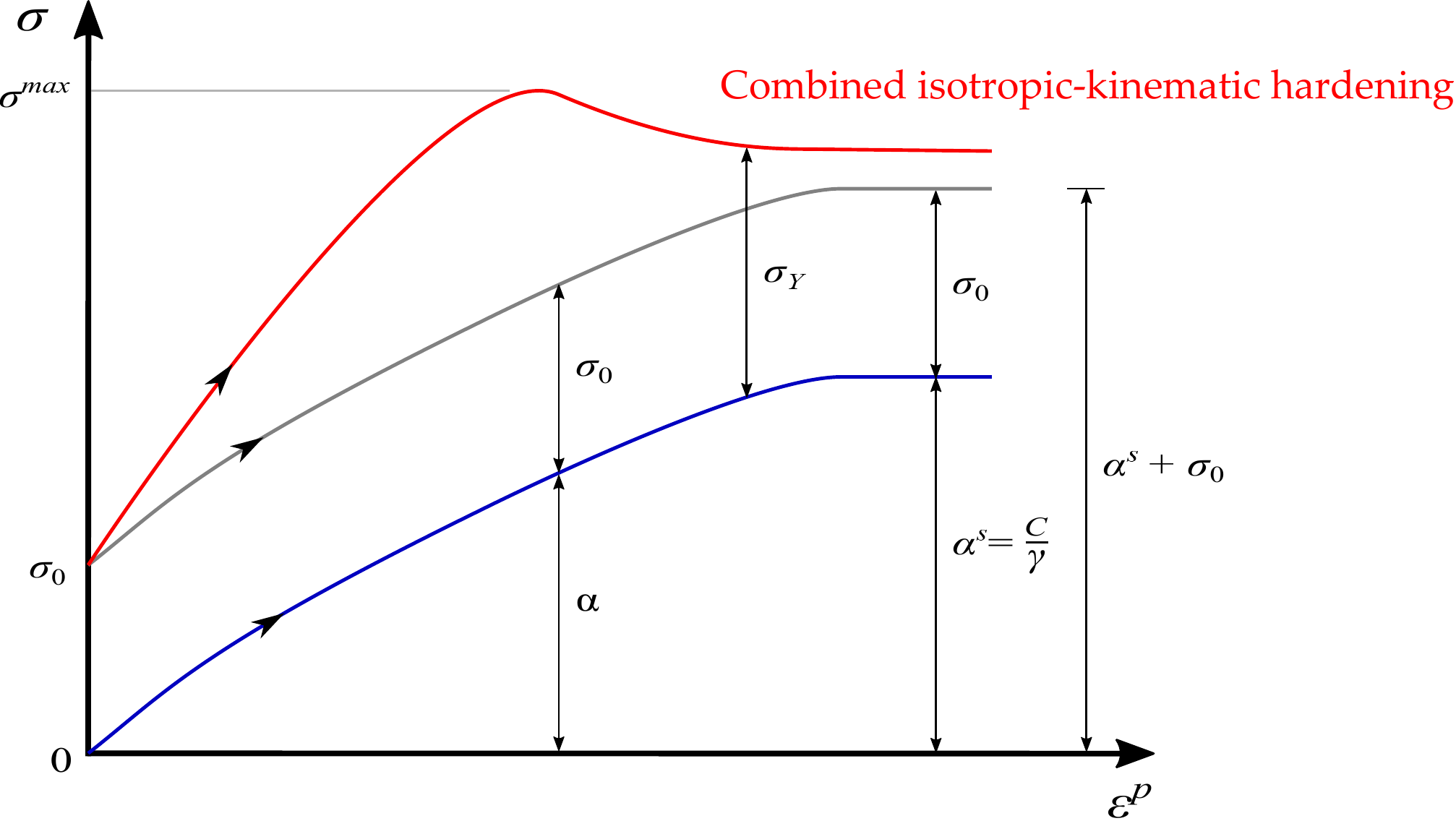}
        \caption{Sketch of the stress versus plastic strain response under uniaxial loading to illustrate the nonlinear combined isotropic/kinematic hardening model.}
        \label{fig:cyclichard}
\end{figure}

The hardening evolution law includes two components: an isotropic hardening one, describing the change in size of the yield surface, and a kinematic hardening one, characterising the translation of the yield surface in the stress space. We employ an exponential isotropic hardening law to relate the change in yield stress to the equivalent plastic strain $\varepsilon^p$ and the initial yield stress $\sigma_0$ as,
\begin{equation}
    \sigma_Y=\sigma_0+Q_\infty \left[ 1- \exp{\left(-b \varepsilon^p \right)} \right] \, ,
\end{equation}

\noindent where $Q_\infty$ is the maximum change in the size of the yield surface and $b$ determines the rate at which the size of the yield surface changes as the plastic strain develops. On the other side, kinematic hardening effects are captured by means of an additive combination of a purely kinematic term, as in Ziegler's linear hardening law \cite{Ziegler1959}, and a relaxation term, which introduces the non-linearity. The model can accommodate several superposed backstresses, which can improve predictions. Accordingly, for each backstress component,
\begin{equation}
    \dot{\bm{\alpha}}_k = \frac{C_k}{\sigma_Y} \left( \bm{\sigma} - \bm{\alpha} \right) \dot{\varepsilon}^p - \gamma_k \bm{\alpha}_k \dot{\varepsilon}^p \,\,\,\,\,\,\,\,\, \text{with} \,\,\,\,\,\,\,\,\, \bm{\alpha} = \sum_{k=1}^{\mathcal{N}} \bm{\alpha}_k \, ,
\end{equation}

\noindent where $C$ and $\gamma$ are parameters calibrated against experimental data from a stabilised stress-strain cycle, and $\mathcal{N}$ is the number of backstresses. Note that the maximum change in backstress is controlled by the ratio $C / \gamma$, where $\gamma$ specifies the rate at which the backstress changes as the plastic strain develops. The model reduces to Ziegler's linear hardening law when $\bm{\alpha}_k=0$ and to an isotropic hardening model when both $C_k$ and $\gamma_k$ are zero. On the other hand, a purely (non-linear) kinematic hardening model is recovered if $\sigma_Y=\sigma_0$. The one-dimensional representation of the non-linear combined isotropic-kinematic hardening law assumed is shown in Fig. \ref{fig:cyclichard}. The maximum uniaxial stress attained is denoted by $\sigma^{max}$ and $\alpha_s$ corresponds to the magnitude of the backstress at saturation. 

\section{Numerical implementation}
\label{Sec:FEM}

Details of the numerical implementation are provided here, starting with the finite element discretisation (Section \ref{sec:FEdis}), followed by the formulation of the residuals and the stiffness matrices (Section \ref{sec:Residuals}), and ending with a description of the quasi-Newton algorithm employed to enable an efficient and robust monolithic implementation (Section \ref{Sec:BFGS}). The theoretical framework outlined in Section \ref{Sec:Theory} is numerically implemented in the commercial finite element package \texttt{ABAQUS} by developing a user-defined \texttt{UELMAT} subroutine, this is described in Section \ref{Sec:ABAQUSdetails}\footnote{The code is available for download at www.empaneda.com/codes.}. 

\subsection{Finite element discretisation}
\label{sec:FEdis}

Making use of Voigt notation, the primal kinematic variables of the problem are discretised in terms of their nodal values $\textbf{u}_i=\{u_x, u_y, u_z\}^T$ and $\phi_i$ at node $i$ as: 
\begin{equation}\label{eq:dicr}
    \textbf{u}=\sum_{i=1}^{m}\bm{N}_i^{\textbf{u}}\textbf{u}_i \text{\hspace{1cm}and\hspace{1cm}}\phi=\sum_{i=1}^{m}N_i\phi_i
\end{equation}

\noindent where $m$ is the total number of nodes per element, $N_i$ is the shape function associated with node $i$ and $\bm{N}_i^{\textbf{u}}$ is the shape function matrix, a diagonal matrix with $N_i$ in the diagonal terms. Making use of the corresponding $\bm{B}$-matrices, which contain the derivatives of the shape functions, the discretised derivatives of \textbf{u} and $\phi$ can be expressed as follows,
\begin{equation}\label{eq:dicr1}
    \pmb{\varepsilon}=\sum_{i=1}^{m}\bm{B}_i^{\textbf{u}}\textbf{u}_i \text{\hspace{1cm}and\hspace{1cm}}\nabla\phi=\sum_{i=1}^{m}\textbf{B}_i^{\phi}\phi_i \, .
\end{equation}

\subsection{Residuals and stiffness matrices}
\label{sec:Residuals}

Let us now formulate the weak form of the problem. Considering the principle of virtual power (\ref{eq:PVW}) and the constitutive choices described in Section \ref{Sec:Theory}, the weak form of the displacement and phase field problems read:
\begin{equation}
  \int_\Omega \Big\{ \big[ {g(\phi)} + \varkappa \big] \bm{\sigma}_0 : \text{sym} \nabla \delta \mathbf{u} - \mathbf{b} \cdot \delta \mathbf{u} \Big\} \, \text{d} V  + \int_{\partial \Omega} \mathbf{T} \cdot \delta \mathbf{u} \, \text{d} S= 0   \, ,
\end{equation}
\begin{equation}
\int_{\Omega} \left\{ {g^{\prime}(\phi)\delta \phi} \, \left( \mathcal{H} + \psi^p \right) +
        \frac{G_c}{2c_w} f \left( \bar{\vartheta} \right)  \left[ \frac{w^{\prime}(\phi)}{2 \ell}  \delta \phi - \ell \nabla \phi  \nabla \delta \phi \right] \right\}  \, \mathrm{d}V = 0  \, .
\end{equation}

\noindent Here, $\bm{\sigma}_0$ is the undamaged stress tensor and $\varkappa$ is a small, positive constant used to prevent ill-conditioning when $\phi=1$; in this work, $\varkappa=1\times10^{-7}$. Also, note that a hybrid approach is used, by which the strain energy density split is only applied to the phase field balance but with $\phi$ degrading the total strain energy density in the linear momentum balance equation \cite{Ambati2015}. Now, introduce the discretisation outlined in (\ref{eq:dicr})-(\ref{eq:dicr1}) into the weak form to derive the corresponding residuals: 
\begin{align}
    & \mathbf{R}_i^\mathbf{u} = \int_\Omega \left\{\left[{g(\phi)}+ \varkappa \right]\left(\bm{B}^\mathbf{u}_i\right)^T \bm{\sigma}_0 - \left( \bm{N}_i^{\mathbf{u}} \right)^T \mathbf{b} \right\} \, \text{d}V - \int_{\partial \Omega}  \left( \bm{N}_i^{\mathbf{u}} \right)^T \mathbf{T} \, \text{d} S \, , \\
    & R_i^\phi = \int_\Omega \left\{ {g^{\prime}(\phi)}N_i \left( \mathcal{H} + \psi^p \right) + \frac{G_c}{2c_w \ell} f \left( \bar{\vartheta} \right) \left[\frac{w^{\prime}(\phi)}{2 } N_i + \ell^2 \,  \left( \mathbf{B}_i \right)^T \nabla\phi\right]\right\}dV \,.
\end{align}

Finally, we obtain the consistent tangent stiffness matrices $\bm{K}$ by differentiating the residuals with respect to the incremental nodal variables as follows:
\begin{align}
    & \bm{K}_{ij}^{\mathbf{u}} = \frac{\partial \bm{R}_{i}^{\bm{u}} }{\partial \bm{u}_{j} } = 
        \int_{\Omega} \left\{ \left[ {g(\phi)}+ \varkappa \right] {(\bm{B}_{i}^{\bm{u}})}^{T} \bm{C}_{ep} \, \bm{B}_{j}^{\bm{u}} \right\} \, \text{d}V \, , \\
    & \bm{K}_{ij}^{\phi} = \frac{\partial R_{i}^{\phi} }{ \partial \phi_{j} } =  \int_{\Omega} \left\{ \left( {g''(\phi)} \left( \mathcal{H} + \psi^p \right) + \frac{G_{c} f \left( \bar{\vartheta} \right)}{4 c_w \ell}  {w''(\phi)} \right) N_{i} N_{j} + f \left( \bar{\vartheta} \right) \frac{G_{c} \ell}{2 c_w}   \, \mathbf{B}_i^T\mathbf{B}_j \right\} \, \text{d}V \, ,
\end{align}

\noindent where $\bm{C}_{ep}$ is the elastic-plastic material Jacobian. The global system of equations then reads,
\begin{equation}\label{eq:intialguess}
{\begin{bmatrix}
\textbf{K}^{\textbf{uu}} & 0\\
0 & \textbf{K}^{\phi\phi}
\end{bmatrix}}\begin{Bmatrix}
\textbf{r}^{\textbf{u}}\\
\textbf{r}^{\phi}
\end{Bmatrix}=\begin{Bmatrix}
\textbf{u}\\
\phi
\end{Bmatrix}
\end{equation}

The $\textbf{u}$ and $\phi$ solutions can be obtained monolithically (simultaneously) or following a staggered (sequential) approach. Staggered solution schemes have been traditionally considered more robust but come at the cost of losing unconditional stability, requiring the use of sufficiently small load increments to ensure that the solution does not deviate from the equilibrium one. This can be a significant bottleneck for phase field fatigue calculations, where errors can accumulate in every cycle. We will show how a robust and efficient monolithic solution scheme can be achieved by using quasi-Newton methods, such as the Broyden-Fletcher-Goldfarb-Shanno (BFGS) algorithm \cite{Wu2020a,TAFM2020}, enabling accurate fatigue crack growth estimations. Note that, a requirement of the BFGS algorithm described below is that the stiffness matrix is symmetric and positive-definite. Hence, $\textbf{K}^{\textbf{u}\phi}=\textbf{K}^{\phi \textbf{u}}=0$.

\subsection{A quasi-Newton solution scheme: the BFGS algorithm}
\label{Sec:BFGS}

Quasi-Newton methods have proven to be very robust when dealing with non-convex minimisation problems (see, e.g. \cite{Li2001,Lewis2013} and Refs. therein). As opposed to standard Newton approaches, the stiffness matrix $\mathbf{K}$ is not updated after each iteration in quasi-Newton algorithms. Instead, an approximation of the stiffness is introduced after a given number of iterations without convergence. This approximated stiffness matrix, $\tilde{\mathbf{K}}$, satisfies:
\begin{equation}
    \tilde{\mathbf{K}}\Delta\mathbf{z}= \Delta\mathbf{r} \,\,\,\,\,\,\,\,\, \text{with} \,\,\,\,\,\,\,\,\,  \mathbf{z}=\begin{Bmatrix} \mathbf{u}\\[0.3em] \bm{\phi}\end{Bmatrix} \, ,
\end{equation}

\noindent Here, $\Delta\mathbf{z}= \mathbf{z}_{t+\Delta t}-\mathbf{z}_t$ and $\Delta\mathbf{r}=\mathbf{r}_{t+\Delta t}-\mathbf{r}_t$. We choose to adopt the BFGS algorithm, in which the approximated stiffness matrix is typically defined as \cite{Geradin1981},
\begin{equation}
    \tilde{\mathbf{K}} = \tilde{\mathbf{K}}_t - \dfrac{(\tilde{\mathbf{K}}_t\Delta\mathbf{z})(\tilde{\mathbf{K}}_t\Delta\mathbf{z}{)}^T}{\Delta\mathbf{z}\tilde{\mathbf{K}}_t\Delta\mathbf{z}}+\dfrac{\Delta\mathbf{r}\Delta\mathbf{r}^T}{\Delta\mathbf{z}^T\Delta\mathbf{r}}
\end{equation}
From a computational perspective, and upon the assumption of a symmetric stifness matrix, the approximated stiffness matrix can be expressed as follows \cite{Matthies1979},
\begin{equation}
       \tilde{\mathbf{K}}^{-1} = \left(\mathbf{I}-\dfrac{\Delta\mathbf{z}\Delta\mathbf{r}^T}{\Delta\mathbf{z}^T\Delta\mathbf{r}} \right)\tilde{\mathbf{K}}_t^{-1}\left(\mathbf{I}-\dfrac{\Delta\mathbf{z}\Delta\mathbf{r}^T}{\Delta\mathbf{z}^T\Delta\mathbf{r}}\right)^{-1}+\dfrac{\Delta\mathbf{z}\Delta\mathbf{z}^T}{\Delta\mathbf{z}^T\Delta\mathbf{r}}  \, .
\end{equation}

\noindent In this way, symmetry and positive definiteness are retained. As it can be observed, if the initial guess is symmetric and positive definite, as in (\ref{eq:intialguess}), the updated $\tilde{\mathbf{K}}$ is also symmetric and positive definite. The BFGS algorithm has been implemented in most commercial finite element packages (such as \texttt{ABAQUS}), often in conjunction with a line search algorithm.

\subsection{Details of the \texttt{ABAQUS} implementation}
\label{Sec:ABAQUSdetails}

A \texttt{UELMAT} subroutine is used to implement the generalised model into the commercial finite element package \texttt{ABAQUS}. The \texttt{UELMAT}, similar to user element (\texttt{UEL}) subroutines, requires defining the element residual and stiffness matrices, such that the coding resembles that of an in-house code (\texttt{ABAQUS} is solely used to assemble the global matrices and solve the system). However, the \texttt{UELMAT} differs from the \texttt{UEL} in that it enables access to the \texttt{ABAQUS}'s material library (\texttt{material\_lib\_mech}). We exploit this to obtain $\bm{\sigma}_0$ and $\bm{C}_{ep}$ for a given strain tensor. In addition, we use the elastic compliance tensor to decompose the total strains into their elastic and plastic counterparts.

\section{Results}
\label{Sec:Results}

We proceed to showcase the capabilities of the computational framework presented by addressing several boundary value problems of particular interest. Firstly, in Section \ref{Sec:Uniaxial}, the number of cycles to failure is estimated under uniaxial cyclic loading and compared with experimental predictions on structural steels. This is accompanied by a parametric study to evaluate the role of the various fracture and fatigue parameters of the model. Secondly, fatigue crack growth in a Compact Tension specimen is predicted, for different solution schemes and phase field models (Section \ref{Sec:CT}). Thirdly, in Section \ref{Sec:AsymmetricNotch}, the interplay between damage and isotropic and kinematic hardening effects is examined in an asymmetrically-notched bar. Finally, we simulate the structural failure of a pipe-to-pipe connection to showcase the capabilities of the model for delivering large scale, 3D predictions (Section \ref{Sec:3Dpipe}).

\subsection{Uniaxial tension-compression experiments}
\label{Sec:Uniaxial}

\begin{figure}[H]
     \centering
        \includegraphics[width=0.7\textwidth]{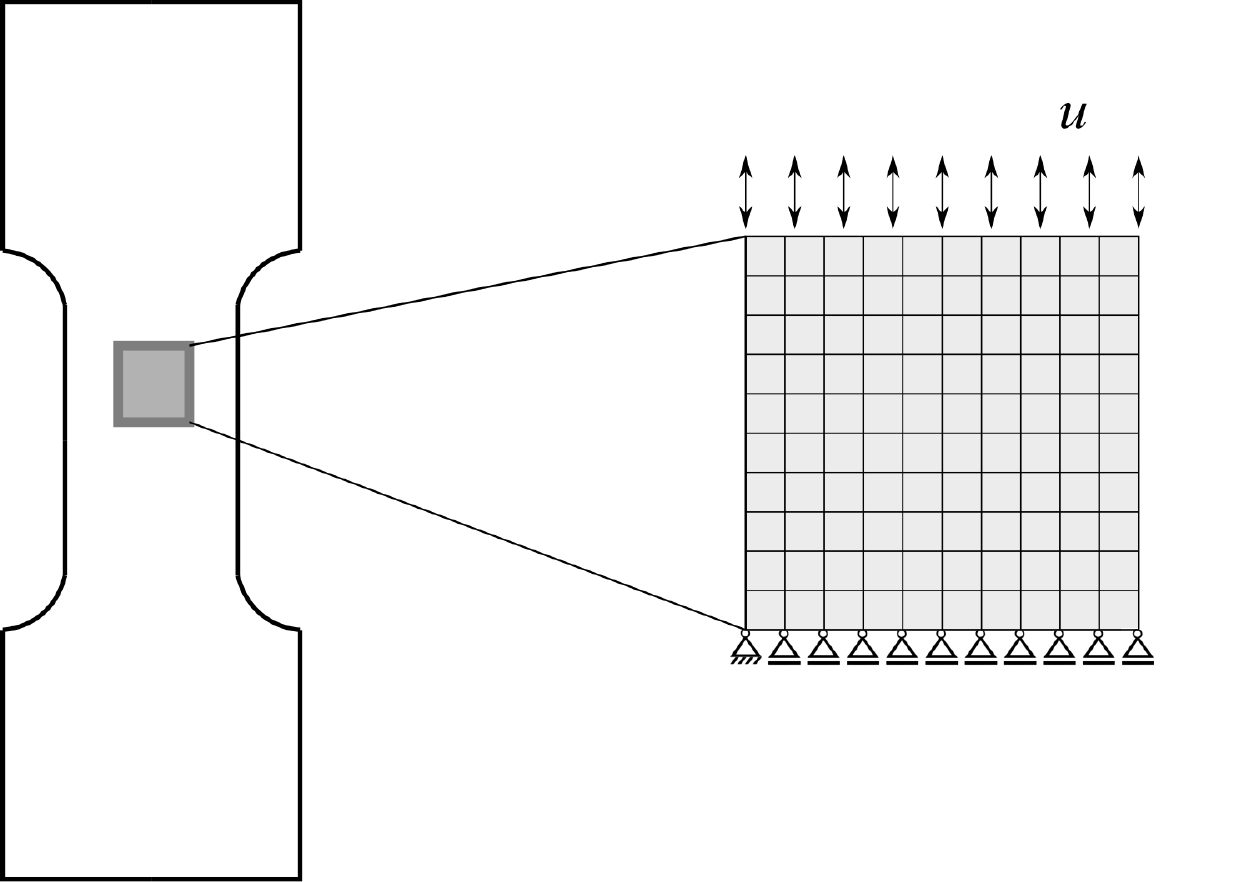}
        \caption{Uniaxial cyclic loading experiments: sketch of a representative piece of material that is subjected to uniform straining; boundary conditions and finite element mesh.}
        \label{fig:model1d}
\end{figure}

\begin{figure}[H]
     \centering
         \centering
         \includegraphics[width=0.7\textwidth]{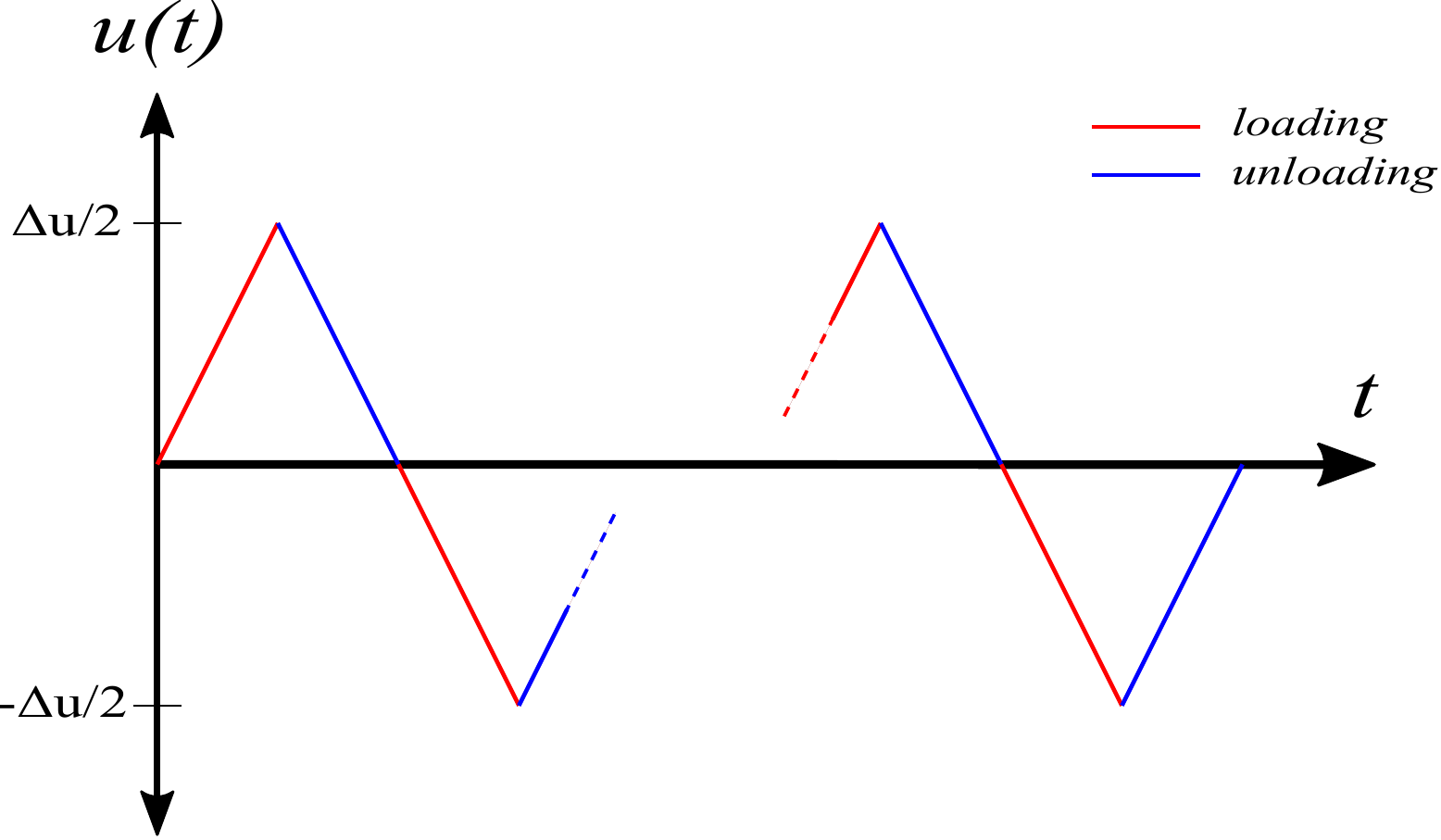}
        \caption{Loading conditions: piece-wise linear variation of the applied displacement under constant amplitude. Red segments correspond to loading stages.}
        \label{fig:u}
\end{figure}

The performance of the proposed modelling framework is first investigated under uniaxial cyclic loading, using strain amplitudes of relevance to low- and extremely low-cycle fatigue applications. Model predictions for various constitutive choices are compared against experimental measurements of number of cycles to failure on hot-rolled structural steels. As the experiments are conducted on smooth, planar samples under uniaxial loading, the boundary value problem is essentially one-dimensional. As illustrated in Fig. \ref{fig:model1d}, calculations are conducted on a square plate with characteristic dimension 1 mm, which is taken to be representative of the conditions experienced by axially loaded test coupons subjected to uniform straining. The square plate is discretised using a uniform finite element mesh, with the characteristic element size being in all cases at least four times smaller than the phase field length scale $\ell$. Linear quadrilateral plane strain elements are used. The vertical displacement is constrained at the bottom of the plate (symmetry conditions) and the loading is applied by prescribing a symmetric quasi-static cyclic displacement with a load ratio of $R=-1$ and a zero mean value, as shown in Fig. \ref{fig:u}. Four strain amplitudes ($\Delta\varepsilon/2$) are considered: 1\%, 3\%, 5\% and 7\%, which are mainly of interest for LCF and ELCF applications.\\

\begin{figure}[H]
     \centering
     \includegraphics[width=1\textwidth]{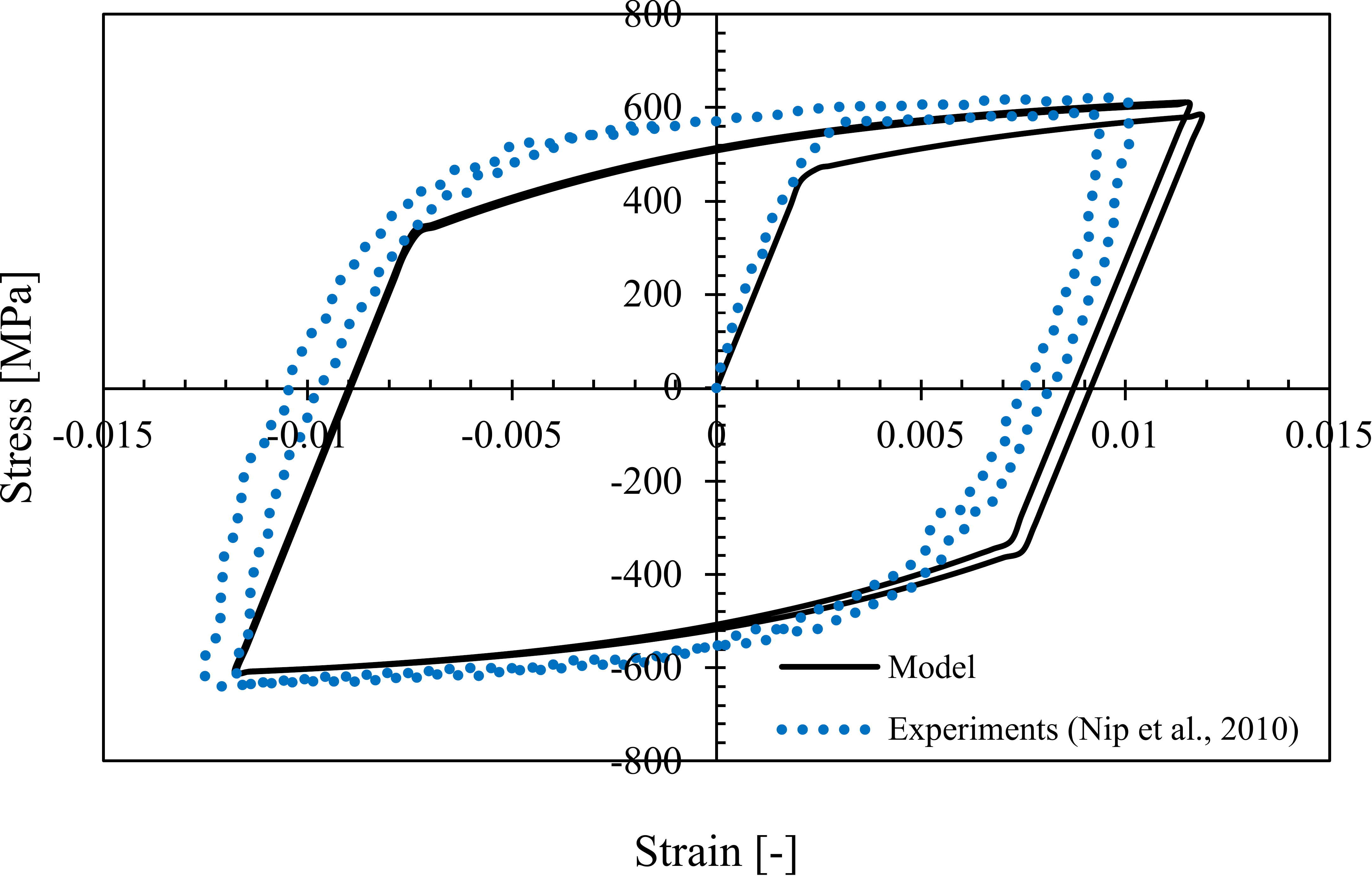}
     \caption{Non-linear kinematic/isotropic hardening material model calibration: model predictions for the first two cycles, compared with the experiments by Nip \textit{et al.} \cite{Nip2010}.}
        \label{fig:ss_model}
\end{figure}

The cyclic plasticity behaviour of the material is calibrated against the experiments by Nip \textit{et al.} on hot-rolled carbon steels \cite{Nip2010,Nip2010a}. The combined non-linear isotropic/kinematic hardening model described in Section \ref{SubSec:Cyclic} must be used to attain a good fit with the experimental data. The magnitudes of the isotropic ($Q_\infty$, $b$) and kinematic hardening parameters ($C$, $\gamma$) that provide the best agreement with the experiments are listed in Table \ref{table:3}, together with the initial yield stress $\sigma_0$ and the elastic properties (Young's modulus $E$ and Poisson's ratio $\nu$). Only one backstress is needed. Fig. \ref{fig:ss_model} shows the agreement between the experimental data \cite{Nip2010} and the present model over the first two cycles, for a representative strain amplitude of 1\%.

\begin{table}[H]
\setlength{\arrayrulewidth}{0.5mm}
\centering
\caption{Material properties that provide the best fit to the experiments by Nip \textit{et al.} \cite{Nip2010,Nip2010a} on a hot-rolled carbon steel exhibiting combined non-linear isotropic/kinematic hardening behaviour.}
\begin{tabularx}{\textwidth}{>{\raggedright\arraybackslash}X >{\raggedright\arraybackslash}X >{\raggedright\arraybackslash}X >{\raggedright\arraybackslash}X >{\raggedright\arraybackslash}X >{\raggedright\arraybackslash}X >{\raggedright\arraybackslash}X >{\raggedright\arraybackslash}X} 
 \hline
E & $\nu$ & $\sigma_0$ & $Q_\infty$ & $b$ & $C$ & $\gamma$   \\ [0.01ex]
[MPa] & [-] & [MPa] & [MPa] & [-] & [MPa] & [-]  \\ [0.5ex]
 \hline
215,960 & 0.3 & 465 & 55 & 2.38 & 23,554 & 139 \\ [1ex] 
 \hline
\end{tabularx}
\label{table:3}
\end{table}

\begin{figure}[H]
     \centering
     \includegraphics[width=1\textwidth]{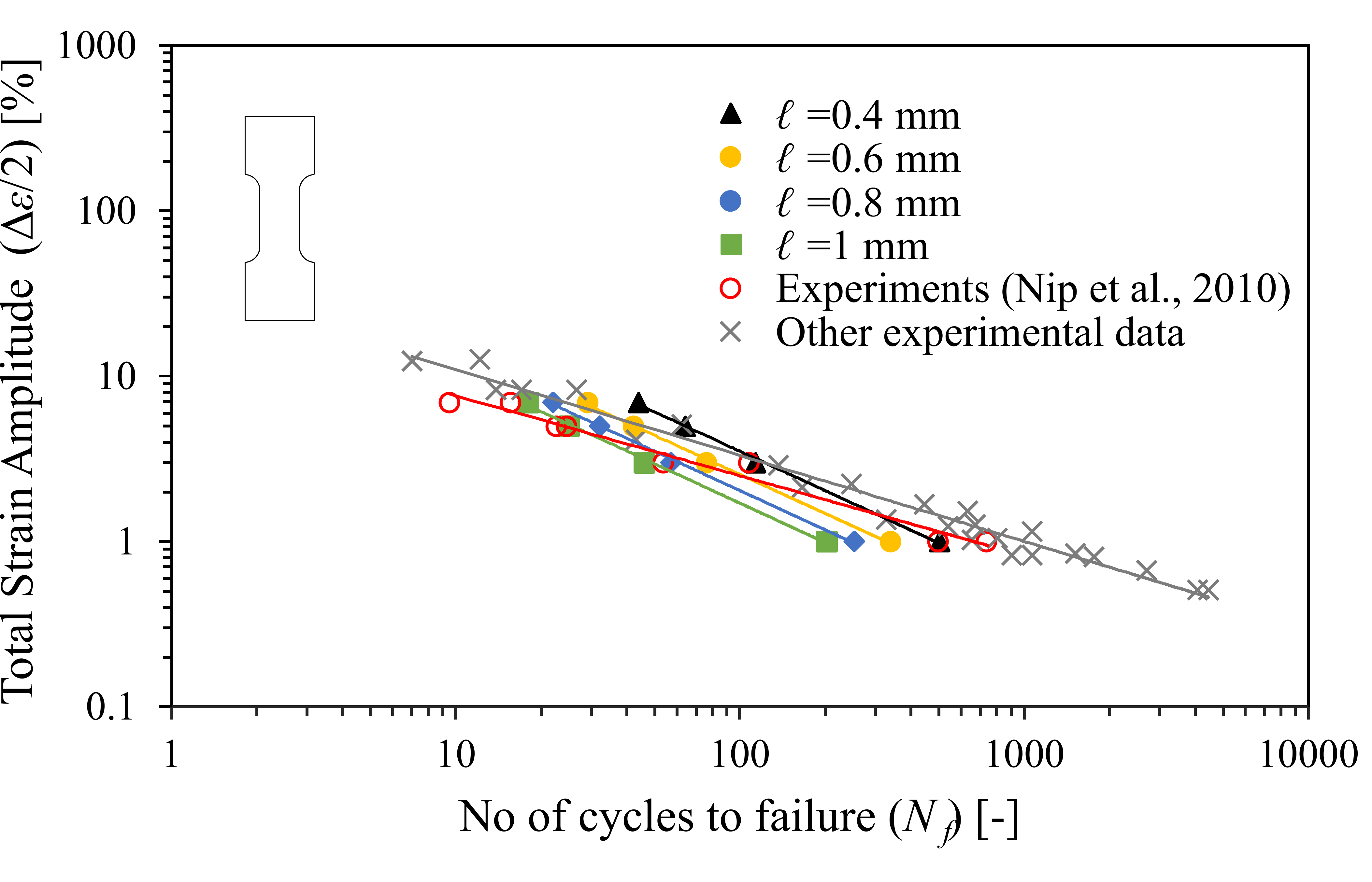}
     \caption{Uniaxial cyclic loading experiments: Total strain amplitude ($\Delta\varepsilon/2$) versus the number of cycles to failure ($N_f$). Numerical results for $G_c=1000$ kJ/m$^2$ and selected values of $\ell$ are compared against experimental data for carbon steels, as reported in the literature \cite{Smith1963,Lefebvre1984,Gong1999,Yang2005b,Nip2010,Nip2010a}.}
        \label{fig:exp}
\end{figure}

The computational results obtained for selected values of the strain amplitude are reported in Fig. \ref{fig:exp}, using a log-log plot. Computations are performed using the \texttt{AT2} model and the asymptotic fatigue degradation function. It is assumed that the sample has failed when the phase field reaches $\phi=0.9$. The magnitude of the material toughness is taken to be equal to $G_c=1000$ kJ/m$^2$ and we vary the phase field length scale $\ell$ to investigate the role of the strength and the fatigue threshold; recall (\ref{eq:AT1AT2strength}) and (\ref{eq:threshold}), respectively. The numerical predictions are shown together with experimental data reported in the literature for carbon steels \cite{Smith1963,Lefebvre1984,Gong1999,Yang2005b} and with the measurements by Nip \textit{et al.} \cite{Nip2010,Nip2010a} on the hot-rolled carbon steel used for calibrating the non-linear combined isotropic/kinematic hardening model. Overall, a good agreement is attained with the experimental data but differences are observed at low and high strain amplitudes. The role of the various fatigue and fracture parameters of the model in changing predictions and improving the agreement with experiments is discussed below. 

\begin{figure}[H]
     \makebox[\textwidth][c]{
     \centering
     \captionsetup[subfigure]{justification=centering}
     \begin{subfigure}[]{0.65\textwidth}
         \centering
         \includegraphics[width=\textwidth]{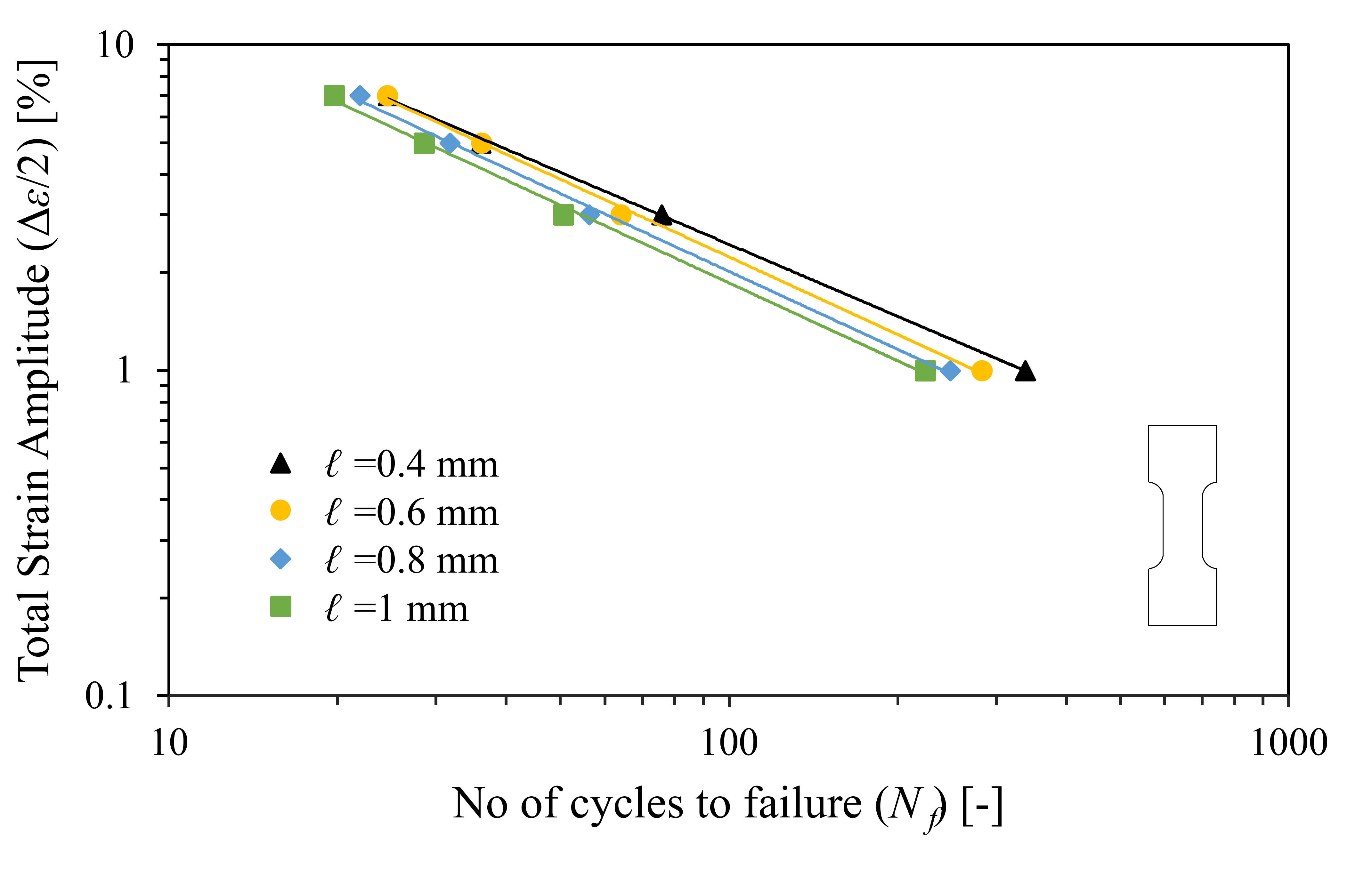}
         \caption{}
     \end{subfigure}
     \hfill
     \begin{subfigure}[]{0.65\textwidth}
         \centering
         \includegraphics[width=\textwidth]{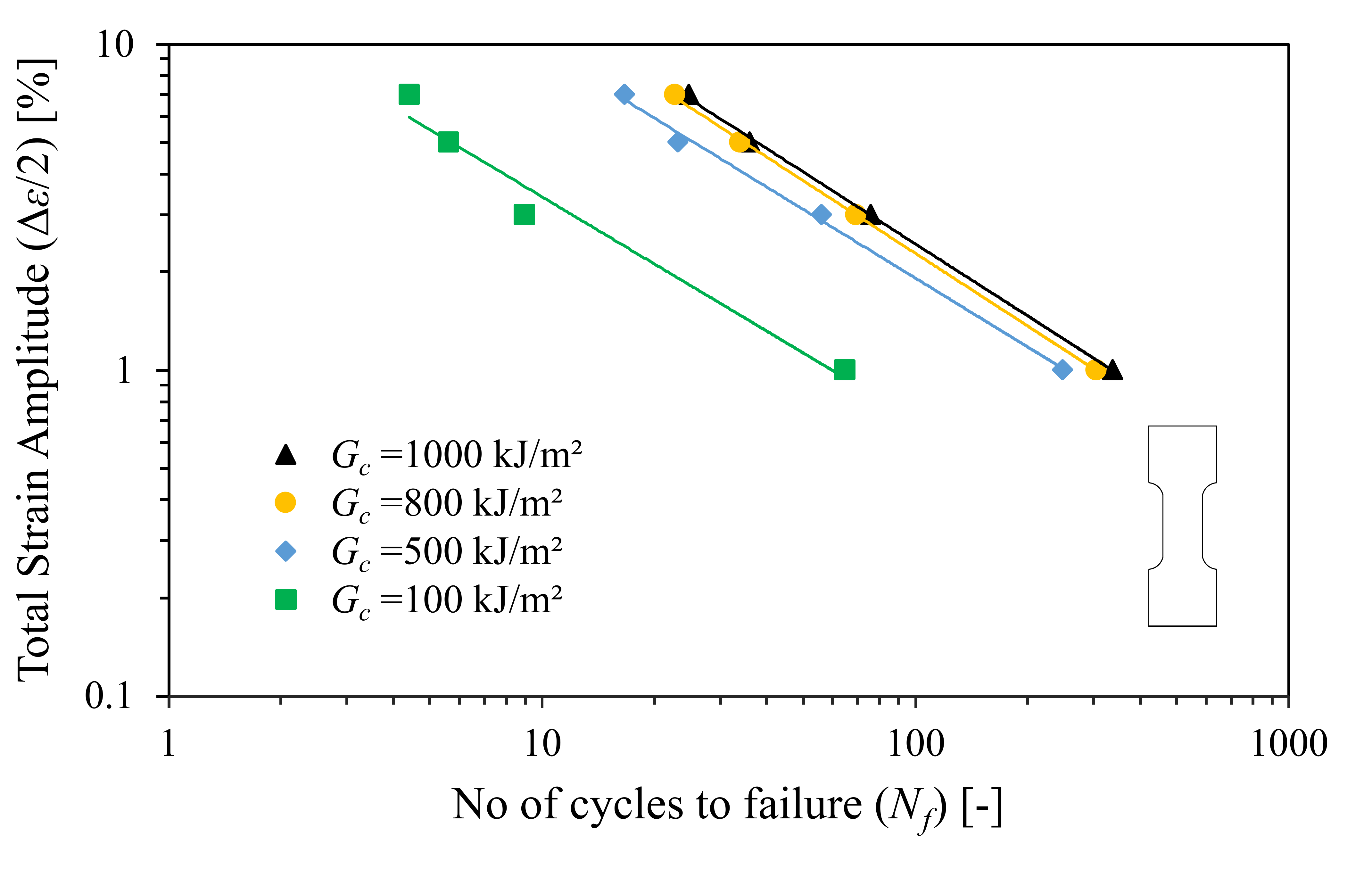}
         \caption{}
     \end{subfigure}}
     \\
     \makebox[\textwidth][c]{
     \begin{subfigure}[]{0.65\textwidth}
        \centering
        \includegraphics[width=\textwidth]{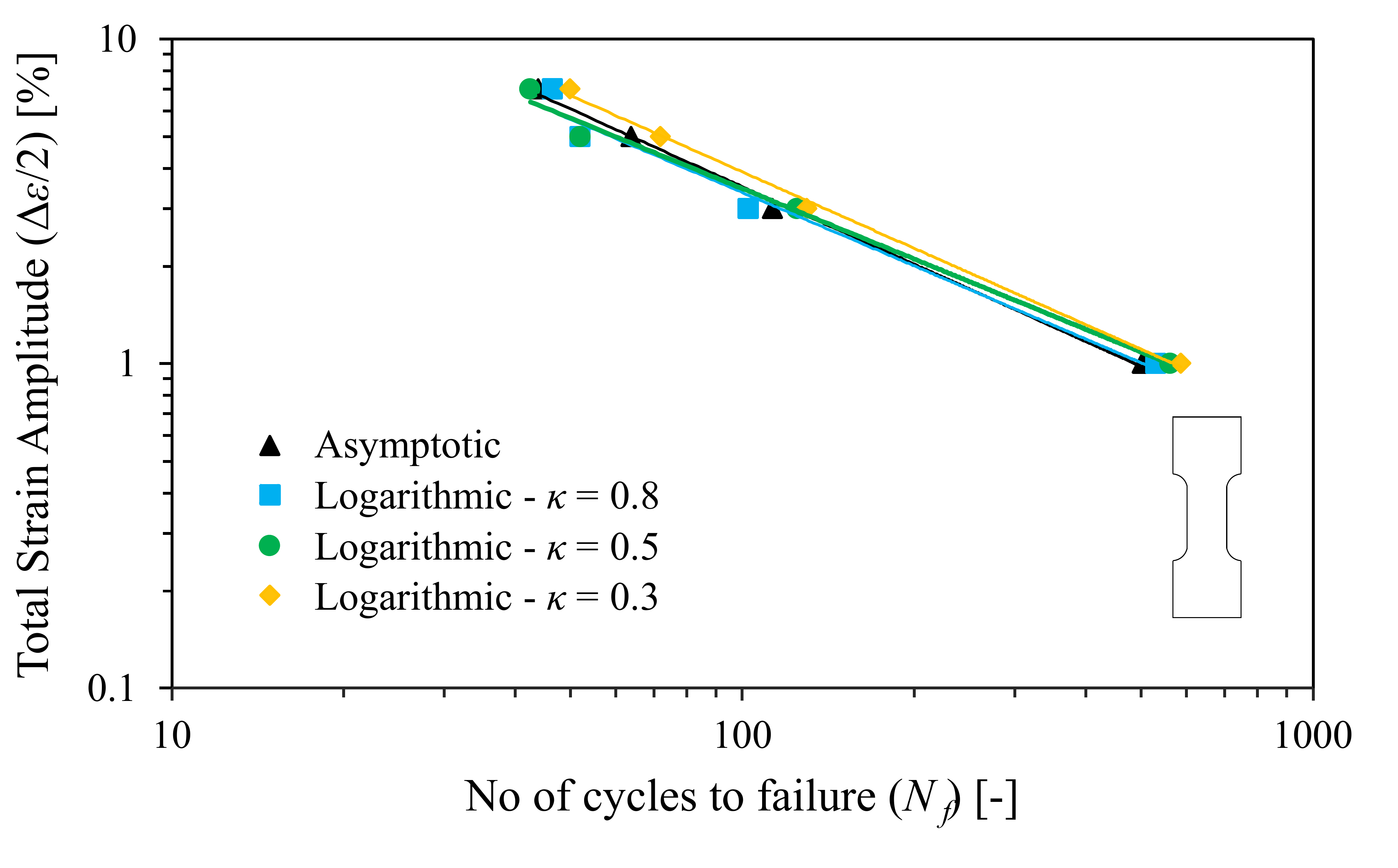}
        \caption{}
        
     \end{subfigure}
     \hfill
     \begin{subfigure}[]{0.65\textwidth}
         \centering
         \includegraphics[width=\textwidth]{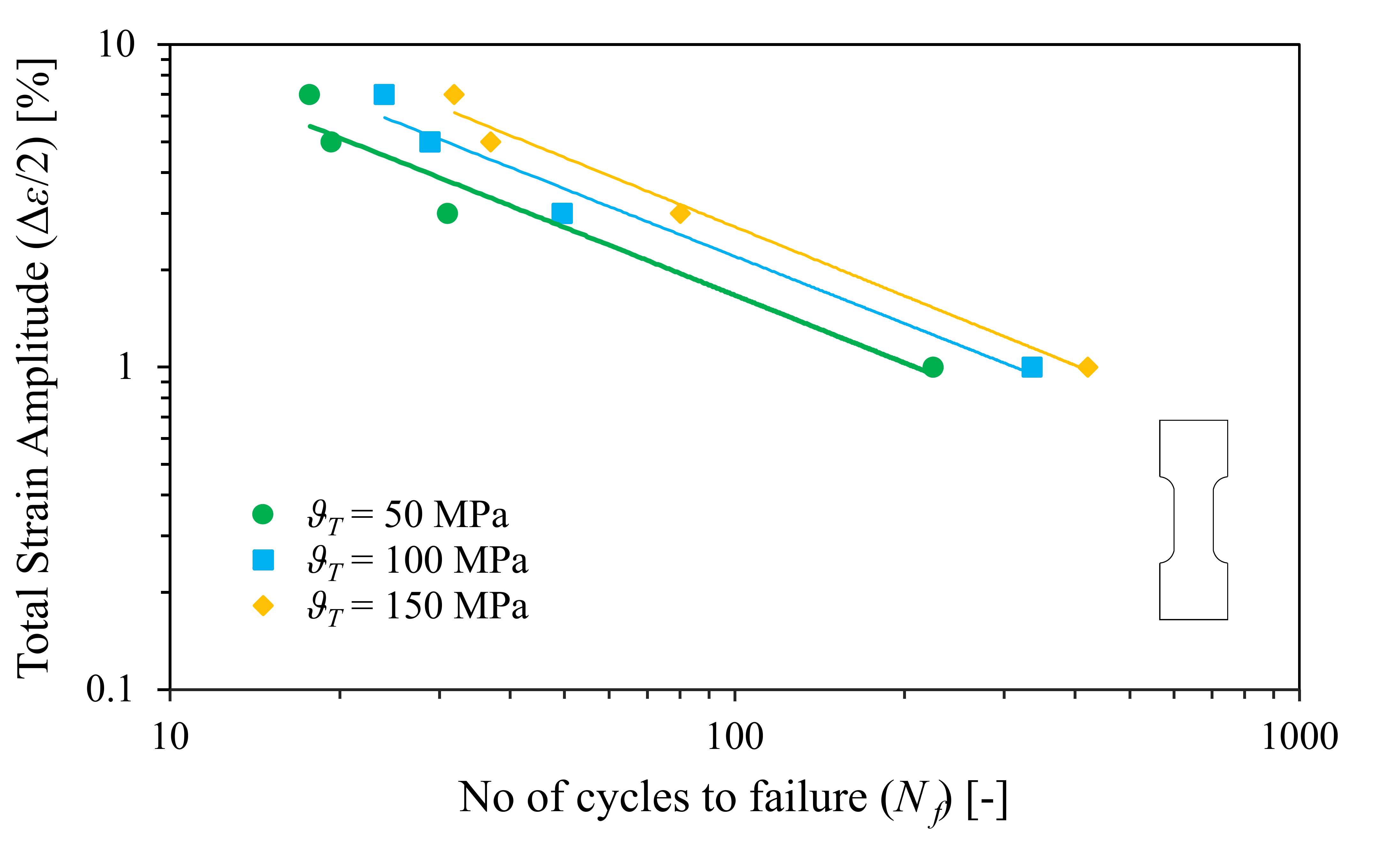}
         \caption{}
     \end{subfigure}}
        \caption{Uniaxial cyclic loading experiments: parametric study. Influence of (a) the length scale $\ell$, (b) the toughness $G_c$, (c) the fatigue degradation function, and (d) the fatigue threshold. In sub-figures (a)-(c), the fatigue threshold $\vartheta_T$ is held fixed, as given by (\ref{eq:threshold}), and in the results shown in (d) have been computed using the asymptotic degradation function.}
        \label{fig:ell}
\end{figure}

A parametric study is conducted and the results are shown in Fig. \ref{fig:ell}. We keep the fatigue threshold $\vartheta_T$ fixed and investigate the influence on the predictions of the phase field length scale, the fracture energy and the fatigue degradation function. Consider first Fig. \ref{fig:ell}a, where the length scale is varied. It can be observed that a smaller length scale translates into a larger fatigue life for a given strain amplitude. This is due to the relation between $\ell$ and the strength in the \texttt{AT2} model - see (\ref{eq:AT1AT2strength})b; the smaller the length scale, the larger the material strength. It can also be noted that the curves are closer to each other, relative to the results shown in Fig. \ref{fig:exp}. The higher sensitivity to $\ell$ observed in Fig. \ref{fig:exp} is due to the relation between the phase field length scale and the fatigue threshold, see (\ref{eq:threshold}). A smaller value of $\ell$ leads to a larger $\vartheta_T$, while in Fig. \ref{fig:ell}a the magnitude of the threshold is fixed at $\vartheta_T=104.167$ MPa (corresponding to $\ell=0.4$ mm and $G_c=500$ kJ/m$^2$) to isolate the influence of the phase field length scale. The influence of the material toughness is evaluated in Fig. \ref{fig:ell}b for $\ell=0.4$ mm. Four values of $G_c$ are considered: 100, 500, 800 and 1000 kJ/m$^2$. The results show significant sensitivity for low $G_c$ values, with the number of cycles to failure increasing with $G_c$, in agreement with expectations. However, one should note that the magnitude of the fracture energy does not typically influence the damage evolution law of cohesive zone models for fatigue \cite{Roe2003,EFM2017}. As for the case of varying $\ell$, changes in $G_c$ do not lead to noticeable variations in the slope of the curve. The role of the fatigue degradation function is explored in Fig. \ref{fig:ell}c for the choices $G_c=$ 1000 $\mathrm{kJ/m^2}$ and $\ell=$ 0.4 mm. Results are obtained for both asymptotic (\ref{eq:fdeg1}) and logarithmic (\ref{eq:fdeg2}) degradation functions, and the latter is assessed for selected values of $\kappa$ (0.3, 0.5, 0.8). Although differences are small for the values considered, it can be observed that the slope of the curve is sensitive to $\kappa$, with larger $\kappa$ values delivering fatigue responses that are more susceptible to changes in the strain amplitude. Thus, the use of a logarithmic degradation function provides additional flexibility, enabling a better match with the experimental results reported in Fig. \ref{fig:exp}. Finally, the role of the fatigue threshold is investigated in Fig. \ref{fig:ell}d. While $\vartheta_T$ can be assumed to depend only on the toughness and the phase field length scale $\ell$, as given in (\ref{eq:threshold}) and adopted in Figs. \ref{fig:exp}a-c, one can also adopt an independent value. The results obtained reveal a longer fatigue life for higher values of $\vartheta_T$, in agreement with expectations.

\subsection{Fatigue crack growth in a Compact-Tension specimen}
\label{Sec:CT}

\begin{figure}[H]
     \centering
        \includegraphics[width=1\textwidth]{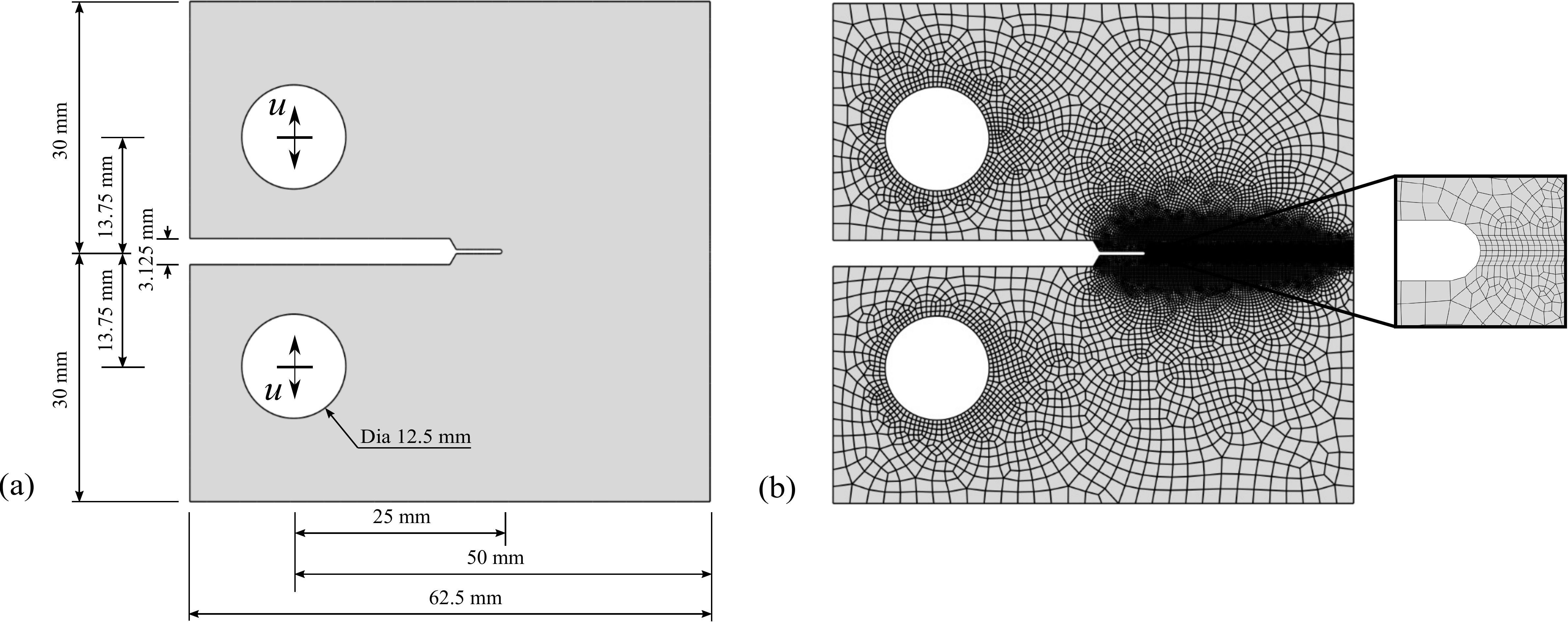}  
        \caption{Fatigue crack growth in a CT specimen: (a) Geometry and boundary conditions and (b) finite element mesh.}
        \label{fig:CT1}
\end{figure}

The influence of the various phase field models and solution schemes presented above is investigated by modelling fatigue crack growth in a Compact Tension (CT) sample. The geometry and dimensions of the sample are shown in Fig. \ref{fig:CT1}, together with the finite element mesh employed. A total of 19,521 4-node plane strain elements are used, with the characteristic element size in the crack propagation region being 5 times smaller than the phase field length scale. The specimen is subjected to symmetric quasi-static cyclic displacement at the pins, with an amplitude of 0.05 mm, a zero mean and a load ratio of $R=-1$. The nonlinear combined isotropic/kinematic hardening material model is used, with the material properties described in Table \ref{table:3}. The assumed values for the toughness and the phase field length scale are $G_c=2.7$ kJ/m$^2$ and $\ell=0.25$ mm, respectively. 

\begin{figure}[H]
     \centering
     \captionsetup[subfigure]{justification=centering}
     \begin{subfigure}[]{0.32\textwidth}
         \centering
         \includegraphics[width=\textwidth]{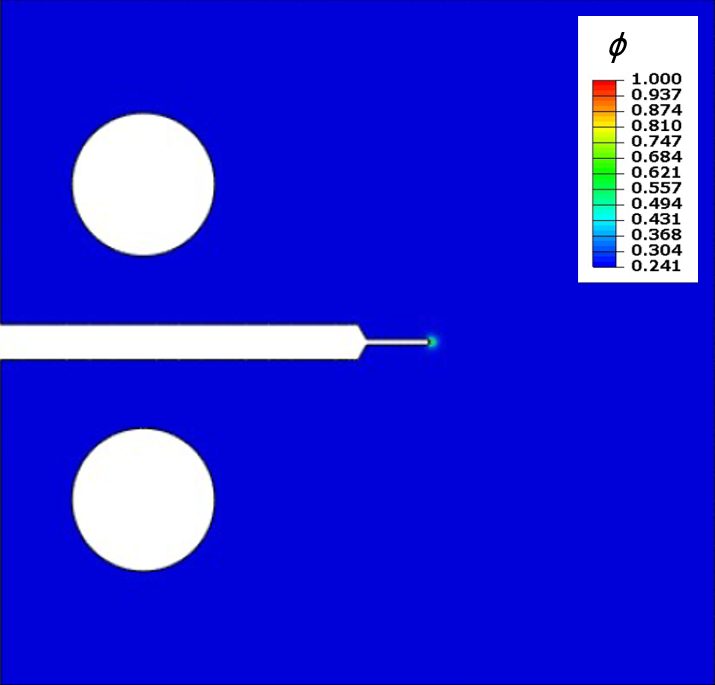}
         \caption{}
         
     \end{subfigure}
     \hfill
     \begin{subfigure}[]{0.32\textwidth}
         \centering
         \includegraphics[width=\textwidth]{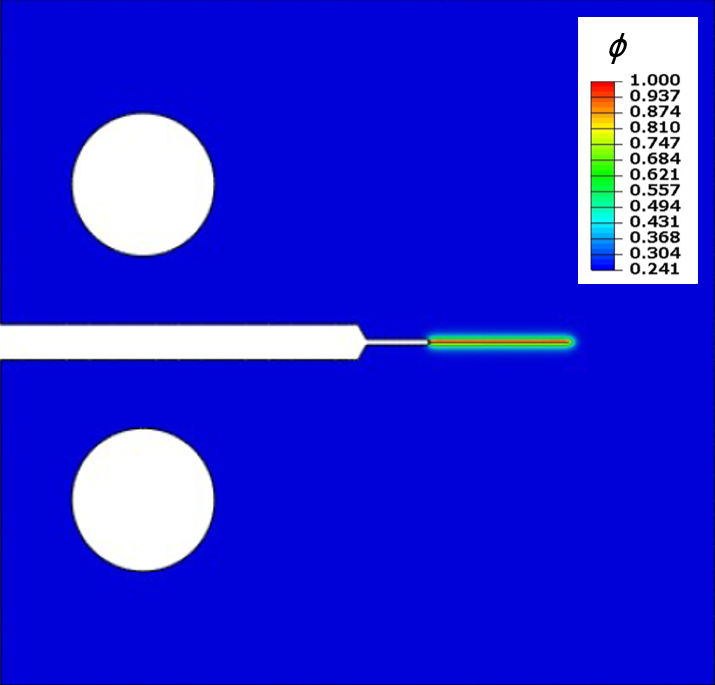}
         \caption{}
         
     \end{subfigure}
     \hfill
     \begin{subfigure}[]{0.32\textwidth}
         \centering
         \includegraphics[width=\textwidth]{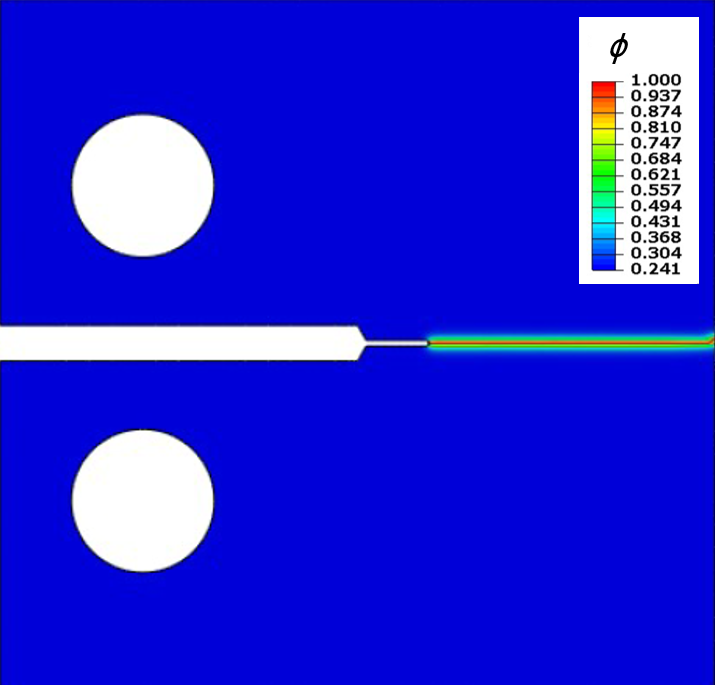}
         \caption{}
         
     \end{subfigure}
        \caption{Fatigue crack growth in a CT specimen. Phase field contours at (a) 12 cycles, (b) 20 cycles and (c) 28 cycles. These representative results have been obtained with the \texttt{AT1} model.}
        \label{fig:CT_AT1}
\end{figure}

\begin{figure}[H]
     \centering
        \includegraphics[width=1\textwidth]{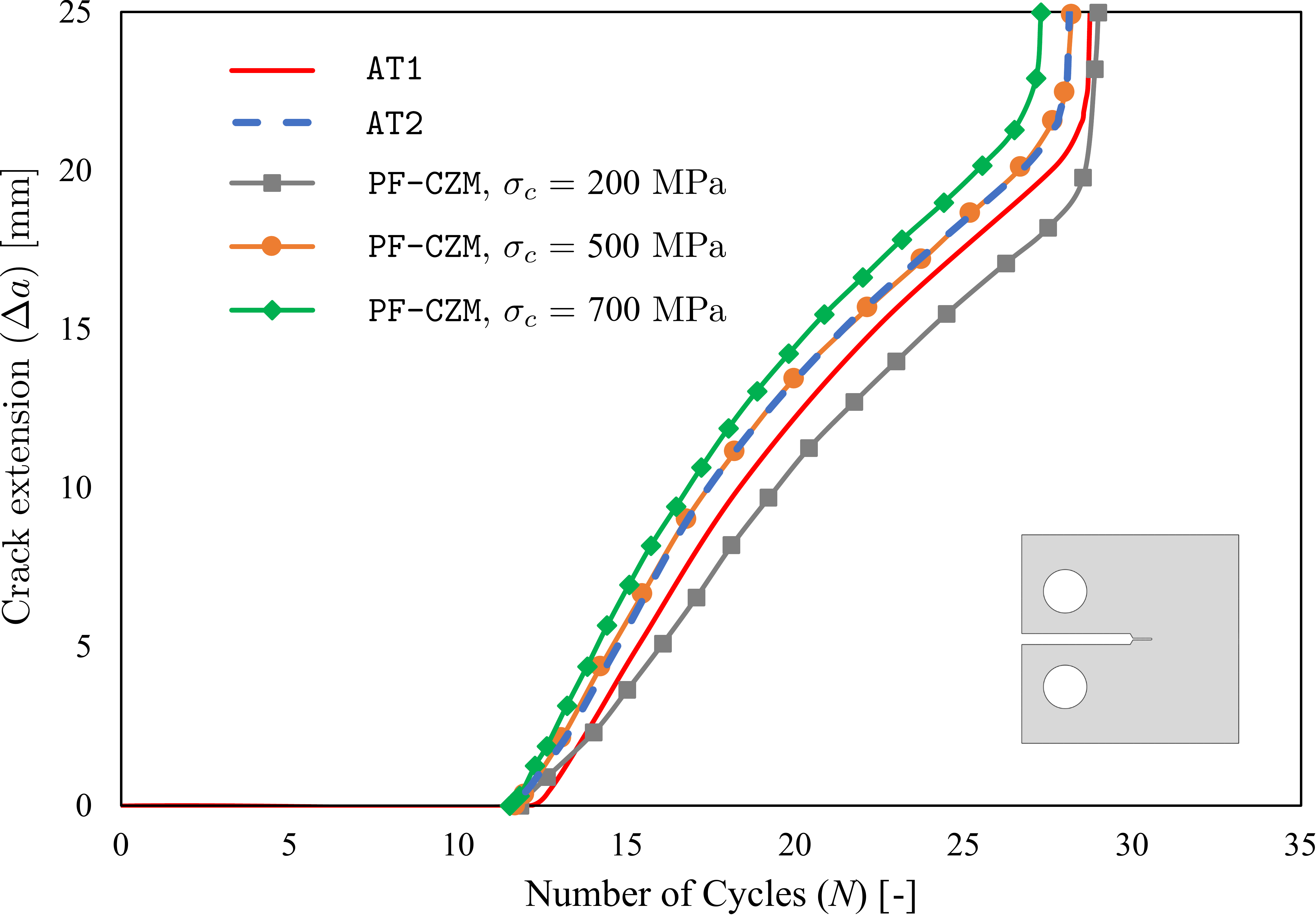}  
        \caption{Fatigue crack growth in a CT specimen: influence of the phase field formulation. Crack extension $\Delta a$ versus number of cycles $N$, as predicted by the \texttt{AT1}, \texttt{AT2}, and \texttt{PF-CZM} models.}
        \label{fig:CT_AT}
\end{figure}

We start by assessing the role of the constitutive choices for the phase field fracture description. As described in Section \ref{Sec:ConstitutiveChoices}, three options are considered, corresponding to the so-called \texttt{AT1}, \texttt{AT2}, and \texttt{PF-CZM} models. In all three cases, the qualitative outcome is in good agreement; as shown in Fig. \ref{fig:CT_AT1}, the crack propagates in a stable manner along the expected mode I trajectory. However, noticeable differences are observed when evaluating the crack extension $\Delta a$ versus number of cycles $N$ curves, see Fig. \ref{fig:CT_AT}.

As shown in Fig. \ref{fig:CT_AT}, fatigue crack growth curves are obtained for \texttt{AT1}, \texttt{AT2}, and \texttt{PF-CZM} models, with three values of the material strength ($\sigma_c$) being used in the \texttt{PF-CZM} case. Consider first the results predicted for \texttt{AT1} and \texttt{AT2} models; similar fatigue crack growth rates are predicted but the number of cycles to failure is higher for the \texttt{AT1} case. This is attributed to the presence of a damage threshold, which is absent in the \texttt{AT2} formulation, and to the higher material strength that results from considering Eq. (\ref{eq:AT1AT2strength}) for the same $G_c$ and $\ell$ values. Specifically, Eq. (\ref{eq:AT1AT2strength}) gives strength values of 935 MPa and 496 MPa for \texttt{AT1} and \texttt{AT2}, respectively. As a result, the initiation of crack growth takes place later for the \texttt{AT1} model, relative to the \texttt{AT2} one. We emphasise that the degradation function is independent of $\ell$ (and thus of the strength) for the \texttt{AT1} and \texttt{AT2} models, see (\ref{eq:gAT1AT2}), and consequently fatigue crack growth rates are similar. However, the \texttt{PF-CZM} model predictions exhibit a different trend. The initiation of crack growth appears to be largely insensitive to the choice of $\sigma_c$ but fatigue crack growth rates differ, with smaller strength values being associated with larger fatigue lives. This is a result of the strength dependency of the phase field degradation function in the \texttt{PF-CZM} model - see (\ref{eq:gPF-CZM}); for a given $\phi$ magnitude, a greater stiffness degradation will be attained for a higher strength. It is also interesting to note that the predictions from the \texttt{AT2} and \texttt{PF-CZM} models coincide when a strength of $\sigma_c=500$ MPa is used in the latter - versus $\sigma_c=496$ MPa for \texttt{AT2}, if estimated from (\ref{eq:AT1AT2strength}). However, the \texttt{AT1} result with an estimated strength of $\sigma_c=935$ MPa lies between the \texttt{PF-CZM} predictions for $\sigma_c=500$ and $\sigma_c=200$ MPa. There is a need to determine the most physically sound relation between $G_c$, $\sigma_c$ and $\ell$, given that the length scale governs the size of the fracture process zone and thus cannot be purely seen as a numerical parameter.

\begin{figure}[H]
     \centering
        \includegraphics[width=1\textwidth]{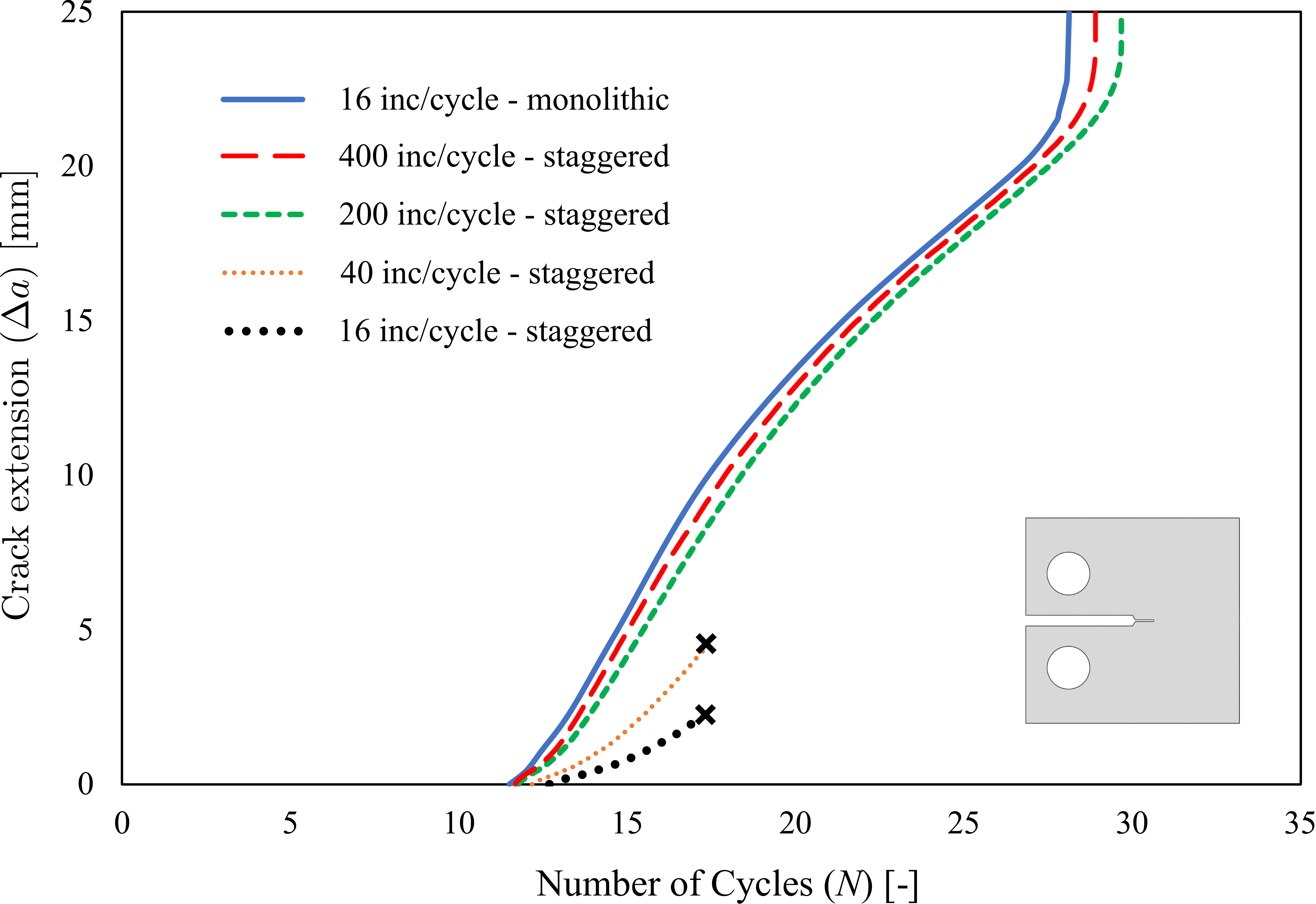}  
        \caption{Fatigue crack growth in a CT specimen: influence of the solution scheme. Crack extension $\Delta a$ versus number of cycles $N$ as predicted with a staggered and a monolithic quasi-Newton scheme. The phase field \texttt{AT2} model has been used.}
        \label{fig:CT_stag}
\end{figure}

The influence of the solution scheme is examined next. Specifically, we compare the performance of the monolithic quasi-Newton algorithm presented in Section \ref{Sec:BFGS} with a single-pass, alternative minimisation staggered scheme.\footnote{We note that while staggered and alternative minimisation schemes are occasionally treated as independent solution strategies in the literature \cite{Storvik2021}, the only difference lies in the inclusion of the $\mathcal{H}$ term. Since it has been reported that enforcing damage irreversibility \textit{via} the history field does not influence the results for these kinds of problems \cite{PTRSA2021}, we here treat both approaches as equivalent.}. Monolithic approaches are unconditionally stable, implying that results are independent of the number of load increments used. This is not the case for staggered approaches, and consequently we obtain results for four load stepping choices: 16, 40, 200 and 400 increments per cycle. As shown in Fig. \ref{fig:CT_stag}, the staggered solution converges slowly towards the monolithic one. Even when using 400 load increments per cycle, the fatigue crack growth curve obtained does not match the monolithic result (obtained with 16 increments per cycle). Differences become very significant when reducing the number of load increments. Furthermore, convergence problems are observed for the simulations with a low number of increments (16 and 40 increments per cycle), with calculations stopping after 17 cycles. Thus, these results show that monolithic quasi-Newton schemes are more robust and significantly more efficient than widely used staggered schemes, which can require prohibitive computation times for accurate cycle-by-cycle predictions of high cycle fatigue. 

\subsection{Failure of an asymmetrically-notched plate}
\label{Sec:AsymmetricNotch}

\begin{figure}[H]
     \centering
        \includegraphics[width=1\textwidth]{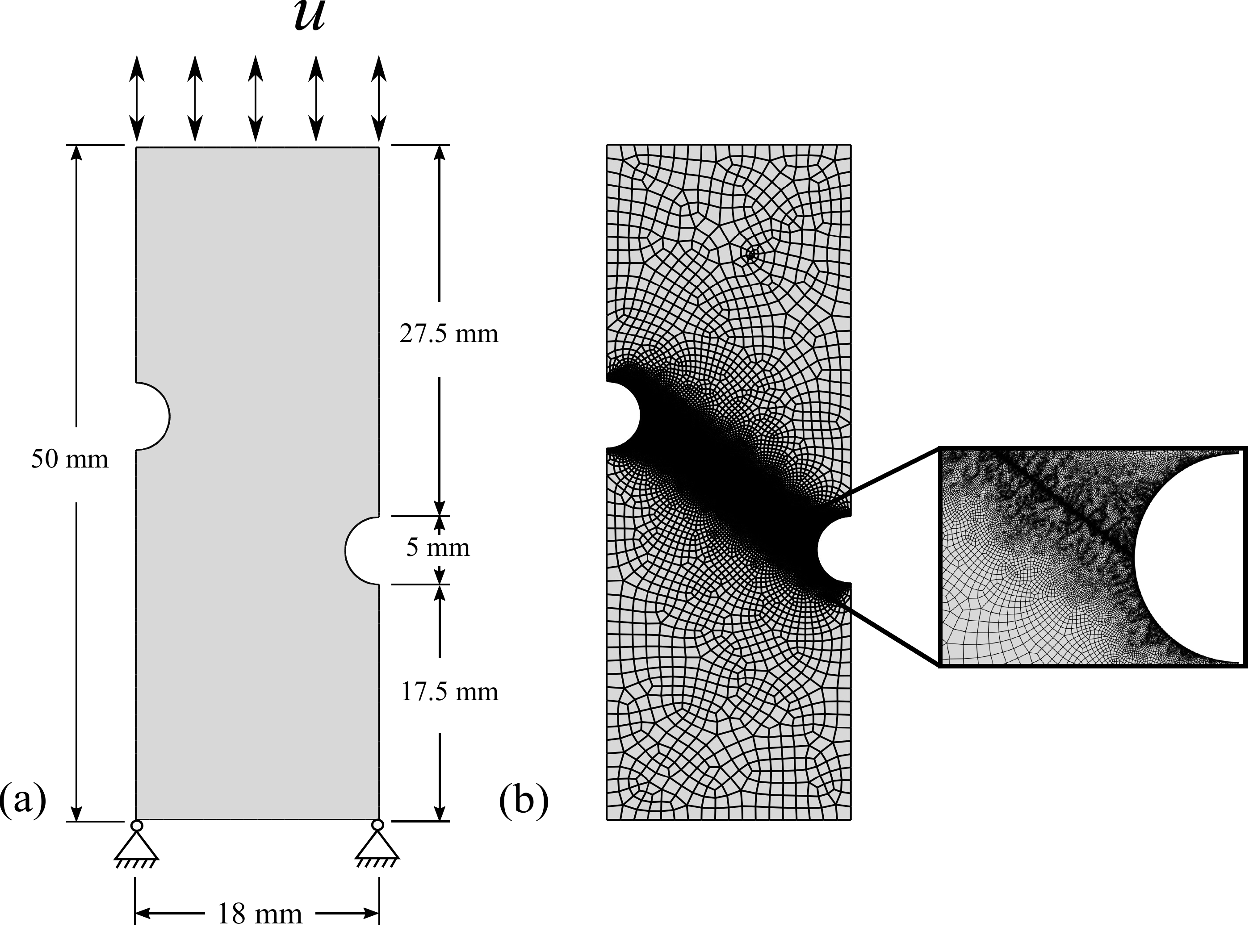}  
        \caption{Asymmetrically-notched specimen subjected to cyclic axial load: (a) Geometry and boundary conditions and (b) finite element mesh.}
        \label{fig:asymm}
\end{figure}

We proceed to investigate the interplay between cyclic hardening and damage by simulating the fatigue failure of an asymmetrically-notched plate. The geometry, boundary conditions and dimensions (in mm) are shown in Fig. \ref{fig:asymm}a. The specimen is discretised with a total of 95,854 linear quadrilateral plane strain elements. As shown in Fig. \ref{fig:asymm}b, the finite element mesh is refined along the crack propagation path, with the characteristic element size being at least four times smaller than the phase field length scale. The specimen is subjected to a symmetric quasi-static cyclic displacement of amplitude 0.5 mm, with zero mean and a load ratio of $R=-1$. The fracture behaviour is characterised by the \texttt{AT2} model, with properties $G_c=8000$ kJ/m$^2$ and $\ell=0.04$ mm.

\begin{figure}[H]
    \centering
     \captionsetup[subfigure]{justification=centering}
     \begin{subfigure}[]{\textwidth}
         \centering
         \includegraphics[]{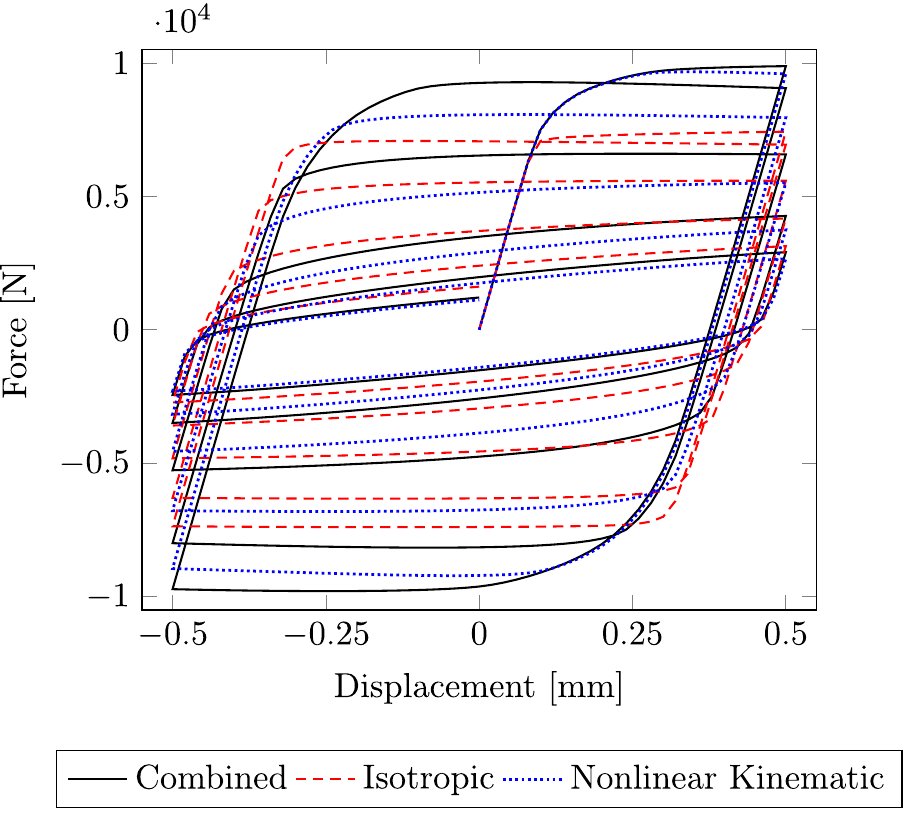}
         \caption{}
         
     \end{subfigure}
     
     \begin{subfigure}[]{\textwidth}
         \centering
         \includegraphics[]{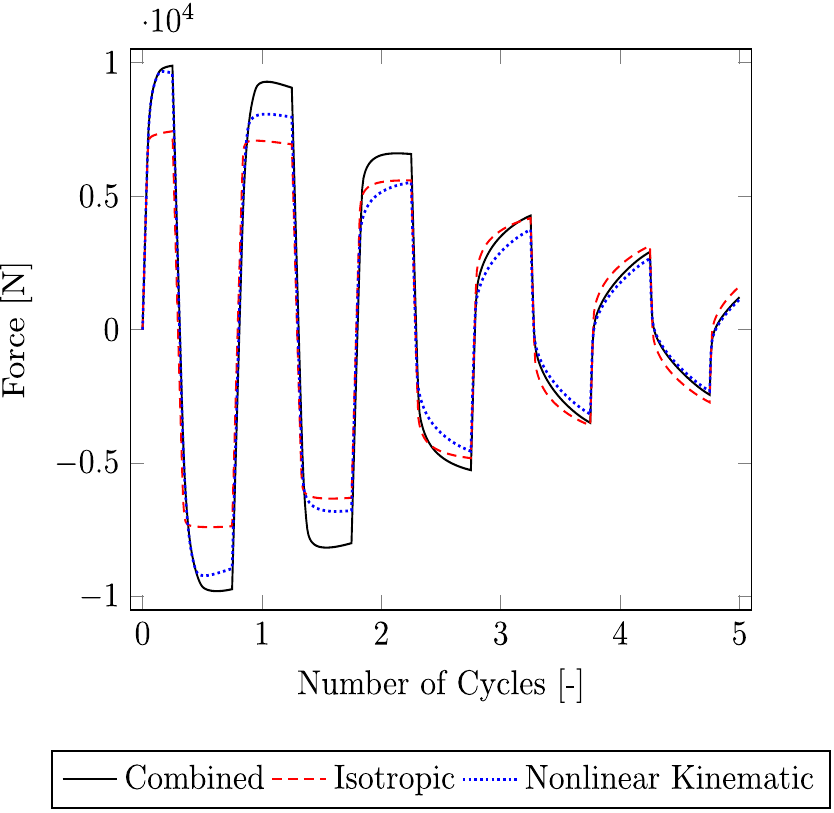}
         \caption{}
         
     \end{subfigure}
    \caption{Asymmetrically-notched specimen subjected to cyclic axial load: (a) Force versus displacement response, and (b) Force versus number of cycles response.}
        \label{fig:fu}
\end{figure}

One of the aims of this case study is to assess the role of kinematic and isotropic hardening effects. Thus, the material constitutive behavior is given by the non-linear combined isotropic/kinematic hardening model described in Section \ref{Sec:ConstitutiveChoices}, with the material properties listed in Table \ref{table:3}, but calculations are also conducted for the cases of purely isotropic hardening ($C=\gamma=0$) and purely kinematic hardening ($\sigma_Y=\sigma_0$). The results obtained for the three material models under consideration are shown in Fig. \ref{fig:fu}, in terms of both the force versus displacement (Fig. \ref{fig:fu}a) and force versus number of cycles (Fig. \ref{fig:fu}b) responses. The results shown in Fig. \ref{fig:fu}a show how the values of maximum force attained change more significantly with the number of cycles when kinematic hardening effects are accounted for, as a result of the Bauschinger effect. Substantial differences are observed over the first cycles, highlighting the importance of properly characterising the cyclic behaviour of the solid. However, the differences between the three models are reduced as damage starts to govern the material response. As shown in Fig. \ref{fig:fu}b, the higher strain energy levels attained with the combined and non-linear kinematic hardening models result in a faster damage rate, while damage is underestimated for the purely isotropic hardening case. This is intrinsic to our choice of an elastic-plastic driving force; see Eqs. (\ref{eq:TotalPotentialEnergy0})-(\ref{eq:StrainEnergyDensity}). 

\subsection{Fatigue crack growth in a pipe-to-pipe connection}
\label{Sec:3Dpipe}

\begin{figure}[H]
     \centering
        \includegraphics[width=1\textwidth]{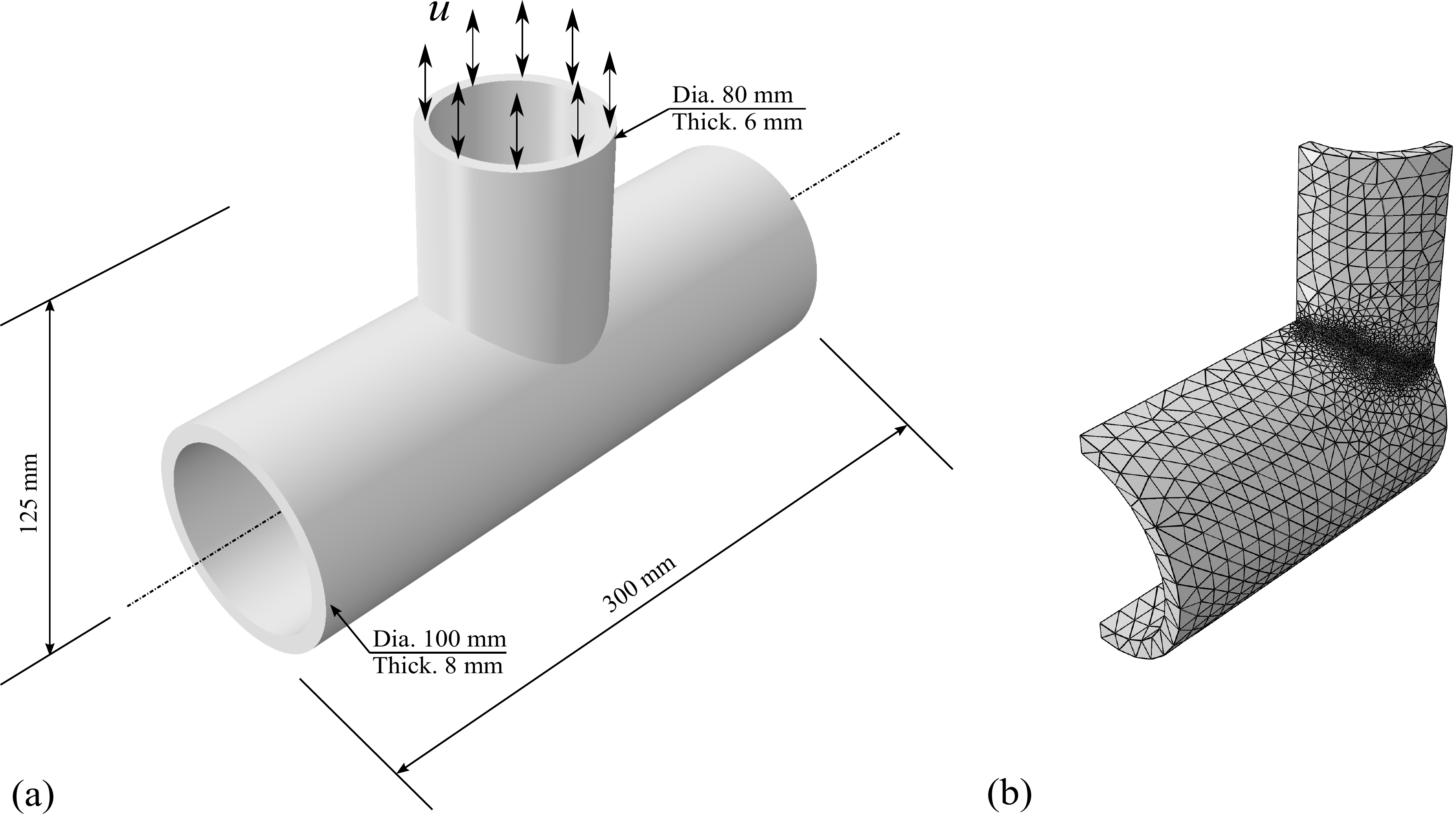}  
        \caption{3D pipe-to-pipe connection subjected to cyclic axial load: (a) Geometry and (b) finite element mesh.}
        \label{fig:3d1}
\end{figure}

Finally, we conclude our numerical experiments by demonstrating the capabilities of the framework presented for predicting large scale fatigue failures in solids exhibiting combined non-linear isotropic/kinematic hardening. To this end, fatigue crack nucleation and growth is simulated in a pipe-to-pipe connection composed of two orthogonal circular pipes - see Fig. \ref{fig:3d1}. The horizontal pipe has an outer diameter of 100 mm and thickness of 8 mm, while the vertical pipe has on outer diameter of 80 mm and thickness of 6 mm. A quarter of the geometry is modelled due to symmetry. The domain is discretised with 10-node tetrahedral solid elements, employing approximately 200,000 degrees-of-freedom. As shown in Fig. \ref{fig:3d1}b, the mesh is refined at the intersection between the two pipes, where fatigue damage is expected to take place. The vertical pipe is subjected to symmetric quasi-static cyclic displacement of amplitude of 0.5 mm with a load ratio of $R=-1$. The horizontal pipeline is clamped on one end. The \texttt{AT-2} phase field damage model is used along with the nonlinear combined isotropic/kinematic hardening constitutive description. The cyclic deformation is characterised by the material properties listed in Table \ref{table:3}. The assumed values of $G_c$ and $\ell$ are $100$ $\mathrm{kJ/m^2}$ and 2.5 mm, respectively.

\begin{figure}[H]
     \centering
     \captionsetup[subfigure]{justification=centering}
     \begin{subfigure}[]{0.4\textwidth}
         \centering
         \includegraphics[width=\textwidth]{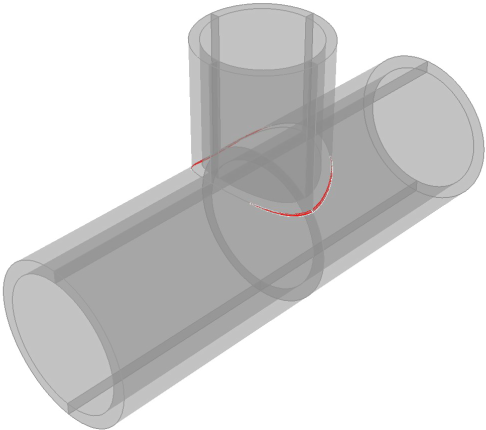}
         \caption{}
     \end{subfigure}
     \hfill
     \begin{subfigure}[]{0.4\textwidth}
         \centering
         \includegraphics[width=\textwidth]{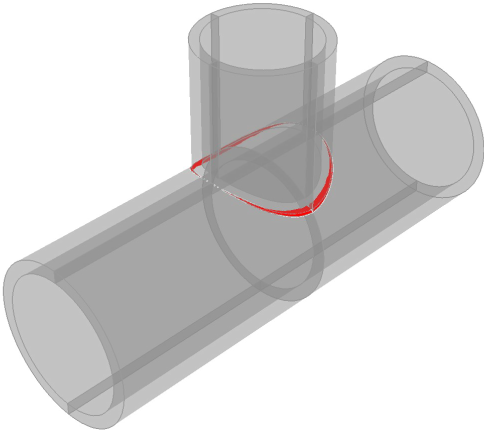}
         \caption{}
     \end{subfigure}
     \\[0.3cm]
     \begin{subfigure}[]{0.4\textwidth}
         \centering
         \includegraphics[width=\textwidth]{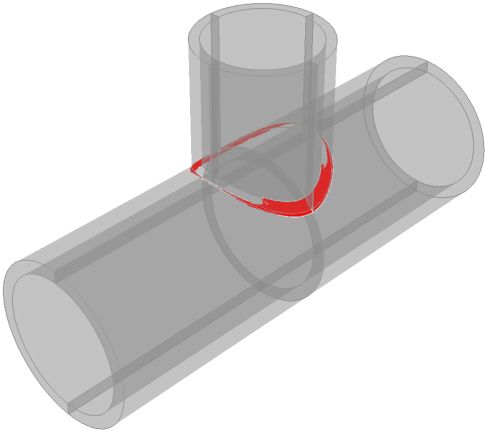}
         \caption{}
     \end{subfigure}
     \hfill
     \begin{subfigure}[]{0.4\textwidth}
         \centering
         \includegraphics[width=\textwidth]{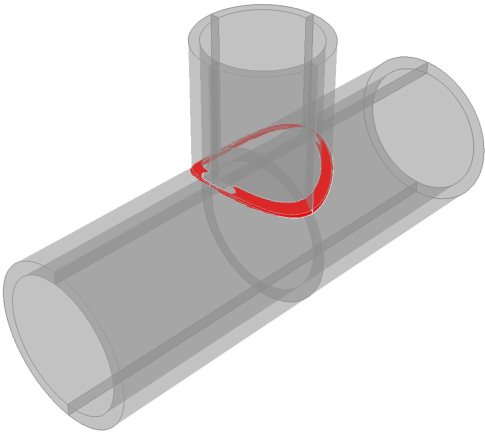}
         \caption{}
     \end{subfigure}
        \caption{3D pipe-to-pipe connection subjected to cyclic axial load. Crack surface ($\phi>0.9$) after: (a) 5 cycles, (b) 8 cycles, (c) 10 cycles and (d) 12 cycles.}
        \label{fig:3d2}
\end{figure}

The process of crack nucleation and growth is depicted in Fig. \ref{fig:3d2}. Four stages of the cracking process are shown, from the initiation of damage to the complete rupture of the pipe-to-pipe connection. We emphasise that no initial defects are introduced in the model. 

\begin{figure}[H]
     \centering
        \includegraphics[width=1\textwidth]{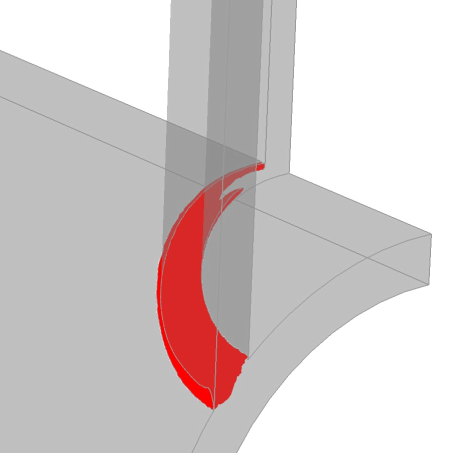}  
        \caption{3D pipe-to-pipe connection subjected to cyclic axial load. Detailed plot of the crack surface ($\phi>0.9$) at 12 cycles.}
        \label{fig:3d3}
\end{figure}

Finally, Fig. \ref{fig:3d3} shows a detailed view of the extent of the crack at the time of complete failure. The results show that the model can naturally predict crack growth due to fatigue damage under cyclic loading in three-dimensional settings for arbitrary geometries and dimensions.

\section{Conclusions}
\label{Sec:Concluding remarks}

We have presented a generalised formulation for modelling fatigue crack growth in elastic-plastic solids. The initiation and growth of cracks is captured by means of a unified phase field formulation, that incorporates as special cases the so-called \texttt{AT1}, \texttt{AT2} and \texttt{PF-CZM} models. Two classes of fatigue degradation functions (asymptotic and logarithmic) are defined to capture fatigue damage, which is driven by both elastic and plastic mechanical fields. The cyclic deformation of the solid is characterised by a combined, non-linear kinematic/isotropic hardening model that can be used to model the cyclic response of a general class of elastic-plastic materials. The theoretical framework presented is numerically implemented using the finite element method and the resulting system of equations is solved in a monolithic manner, by exploiting a quasi-Newton algorithm. Several 2D and 3D case studies are modelled to gain insight into the role of the phase field formulation and the interplay between damage and cyclic hardening effects. Key findings include:\\

\noindent (i) The model provides a good agreement with experimental data on carbon steels. However, the additional flexibility provided by the logarithmic degradation function is needed to capture the appropriate damage scaling. This comes at the cost of defining one additional parameter.\\

\noindent (ii) Fatigue crack growth predictions differ between those obtained with phase field models based on the Ambrosio-Tortorelli functional (\texttt{AT1}, \texttt{AT2}) and those inspired in cohesive laws, such as the \texttt{PF-CZM}. In the \texttt{PF-CZM}, higher fatigue crack growth rates are predicted with increasing material strength $\sigma_c$, due to the $\sigma_c$-dependency of the fracture degradation function.\\

\noindent (iii) The use of quasi-Newton monolithic solution schemes delivers a more robust and efficient performance than widely-used staggered approaches. The staggered scheme requires over 25 times more load increments to match the monolithic result. The implications are important, given the inherent cost of cycle-by-cycle predictions.\\

\noindent (iv) For a crack driving force that includes both elastic and plastic contributions, neglecting kinematic hardening phenomena such as the Bauschinger effect is non-conservative, and can lead to a significant underestimation of damage.\\

The case studies addressed show how the modelling framework presented can be used to efficiently predict fatigue crack initiation and growth in arbitrary dimensions and geometries, and for solids exhibiting complex cyclic deformation responses. The finite element code developed can be downloaded from www.empaneda.com/codes.

\section{Acknowledgments}
\label{Acknowledge of funding}

E. Mart\'{\i}nez-Pa\~neda acknowledges financial support from the EPSRC (grant EP/V009680/1) and from the Royal Commission for the 1851 Exhibition (RF496/2018). Z. Khalil acknowledges the MSc Scholarship support provided by the Department of Civil and Environmental Engineering at Imperial College London.




\bibliographystyle{elsarticle-num}
\bibliography{library}

\end{document}